\documentclass[twocolumn]{revtex4-1}

\pdfoutput=1

\usepackage{subcaption}
\usepackage{hyperref}
\usepackage{enumerate}
\usepackage{booktabs} 
\usepackage{color}
\usepackage{amsfonts}
\usepackage{amsmath}
\usepackage{graphicx}
\usepackage[ruled]{algorithm2e}

\usepackage{diagbox}
\usepackage{mathtools}
\mathtoolsset{showonlyrefs}

\usepackage{enumerate}
\usepackage{amssymb}
\usepackage{amsthm}

\newcommand*{\trace}{\mathsf{Tr}} 
\newcommand\CoAuthorMark{\footnotemark[\arabic{footnote}]}
\begin{document}

\author{Moritz August\footnote{These two authors contributed equally.}}
\thanks{These two authors contributed equally. This work was partly funded by the \emph{Elite Network of Bavaria} via the doctoral programme \emph{Exploring Quantum Matter}.}
\affiliation{Department of Informatics, Technical University of Munich, 85748 Garching, Germany (august@in.tum.de)}
\author{Xiaotong Ni\protect\CoAuthorMark}
\thanks{These two authors contributed equally. This work was partly funded by the \emph{Elite Network of Bavaria} via the doctoral programme \emph{Exploring Quantum Matter}.}
\affiliation{Max-Planck Institute for Quantum Optics, 85748 Garching, Germany (xiaotong.ni@mpq.mpg.de)}

\title{Using Recurrent Neural Networks to Optimize Dynamical Decoupling for Quantum Memory}

\begin{abstract}
We utilize machine learning models which are based on recurrent neural networks to optimize dynamical decoupling (DD) sequences. DD is a relatively simple technique for suppressing the errors in quantum memory for certain noise models. In numerical simulations, we show that with minimum use of prior knowledge and starting from random sequences, the models are able to improve over time and eventually output DD-sequences with performance better than that of the well known DD-families. Furthermore, our algorithm is easy to implement in experiments to find solutions tailored to the specific hardware, as it treats the figure of merit as a black box.

\end{abstract}

\maketitle

\section{Introduction}
A major challenge of quantum information processing (e.g. quantum computation and communication) is to preserve the coherence of quantum states. While in principle we can build a fault-tolerant quantum memory or universal quantum computer once the error rate of the device is below a certain threshold, it is still beyond nowadays experimental capacity to build a decent size quantum computer. One less explored area is the optimization of implementing a fault-tolerant protocol on a concrete experimental setting. This is often a tedious problem, due to the amount of details in the real devices, and the fact that the architectures of both experimental devices and theoretical protocols are still rapidly changing. Thus, an attractive approach is to automatize this optimization task. Apart from convenience, it is conceivable that with less human intuition imposed, the upper bound of the performance will be higher. This has previously been proven to be true in fields such as computer vision where artificial neural network (ANN) models that try to solve tasks without using hand-crafted representations of data have overtaken approaches based on human insight in tasks like image classification and object recognition~\cite{krizhevsky2012}. Another interesting recent example is the ability of ANNs to learn how to play games on a human or even super-human level without any or just little prior knowledge about the respective games~\cite{mnih2013, silver2016}.

Automatically optimizing parameters in real (or numerical simulations of) experiments is not a new idea.
For example, it has been applied to optimizing the pulse shape of a laser, the parameters of Hamiltonians to achieve certain unitary operations or parameters of dynamical decoupling and cold atom experiments.
Most works that attempt to obtain optimal parameters use genetic algorithms~\cite{Judson1992,Heras2015, Geisel2013,Pawela2016} (and to some degree~\cite{Krenn2016}) or local searches such as gradient descent~\cite{Chen2013,Dong2014,Pawela2016,Banchi2015,machnes2011comparing} and the Nelder-Mead simplex method~\cite{Biercuk2009,Doria2011,Kelly2014}.
It is argued that by using these optimization methods directly on the experiments, we can avoid the hardness of modelling the imperfect control and the system-environment interaction.
However, one possible weakness of these optimization methods is that they generate new trials only by looking at a fixed number of previous ones, and often they need to restart once they reach a local minimum.
Thus, in the long run, they do not fully utilize all the data generated by the experiments.
Conceptually, the works listed above resemble more moutain climbing than the learning processes that we have in mind as physicists.


In this work, we propose an orthogonal approach, where we try to mimic the structure of good parameters by building a model that approximates the probability distribution of these parameters. After an initial optimization, this model can then be used to efficiently generate new possible trials and can be continuously updated based on new data. 
In particular, based on the problem we attempt to solve, we choose this model to be a variant of the recurrent neural network (RNN), which makes our approach very similar to the way in which natural languages or handwriting are currently modelled.
This ansatz enables us to exploit the models and insights developed by the machine learning community and possibly translate further progress there into the field of quantum control.
It is worth pointing out that the machine learning part of this work is purely classical; only the (classical) data are related to quantum time evolution.
Among the previous work, the approach in~\cite{Wigley2015} is the most similar to ours, as they attempt to build a model from the data and utilize the model to perform optimization.
(Classical) machine learning is also used in~\cite{Combes2014,Orsucci2015,Tiersch2015} to characterize the error models in quantum error correction and to react accordingly.

To demonstrate the feasibility of using our method to help optimizing quantum memory, we consider the problem of automatically learning and optimizing dynamical decoupling sequences (almost) without using any prior knowledge.
Dynamical decoupling (DD)~\cite{Viola1999} is a technique which combats certain noise by applying a sequence of unitary operations on the system (see~\cite{Souza2011,Quiroz2013} for a review). It has a less stringent requirement compared to general error correction protocols, which allows it to be demonstrated in experiments~\cite{Biercuk2009,DeLange2010,Souza2011} in contrast to other methods. Moreover, known classes of good DD-sequences have a relatively simple and well-defined structure. Based on the assumption that this holds true also for yet unknown and possibly better classes of sequences, it is conceivable that a learning algorithm could eventually sample them without the need of using heavy mathematics.

To clarify, we do \textbf{NOT} attempt to solve the following questions
\begin{itemize}
	\item \textit{What do RNNs try to learn?} It is known that RNNs can incorporate both short and long range correlation, which is desirable in our case, but it is unclear which one the gradient training method prioritizes. Indeed, it is an ongoing study to understand the behavior of RNNs~\cite{karpathy2015}. Nevertheless, we choose to use RNNs since there are heuristic arguments on the advantage of them compared to similar models~\cite{lipton2015critical}, and they benefit a lot from modern machine learning libraries and hardwares.
	\item \textit{What is the optimal machine learning algorithm to find the best DD sequences?} It is clear that we cannot claim our algorithm is the best one as there is not much theoretical understanding on RNNs. Indeed, the authors believe there is much room for improvement, possibly by using better heuristics or take into account more prior knowledge of DD. However, our work demonstrates that with a general model and a small amount of human effort, we can already achieve non-trivial results for certain problems.
\end{itemize}

\section{Background}

\subsection{Dynamical decoupling}
\label{sec_dynamical_decoupling}

The majority of dynamical decoupling schemes are designed for error models where the system-environment interaction can be described by a Hamiltonian. We will use $\mathcal{H}_S$ and $\mathcal{H}_B$ to denote the Hilbert space of the system and environment (often called bath), respectively. The difference between system and environment is that the former represents the part of the Hilbert space we can apply the Hamiltonian on and in which we store quantum information. The total noise Hamiltonian is
\begin{align}
H_0=H_S\otimes I_B + I_S \otimes H_B+H_{SB}.
\end{align}
Without intervention, in general $H_0$ would eventually destroy the quantum states we store on $\mathcal{H}_S$. To suppress this noise, we could apply a time dependent Hamiltonian $H_C(t)$ to the system, which makes the total Hamiltonian $H(t)=H_0+H_C(t)$.
In the ideal case, we can control $H_C(t)$ perfectly and reach very high strength (i.e. norm of the Hamiltonian), which allows the ideal pulse
\begin{align*}
V(t)=O\delta (t-t_0).
\end{align*}
It applies a unitary operator $e^{-iO}$ to the system for an infinitely small duration (we set $\hbar=1$ in this work).
A very simple DD-scheme for a qubit (a two level system $S$) is the $XY_4$ sequence: it applies pulses of the Pauli-matrices $X$ and $Y$ alternatingly with equal time interval $\tau_d$ in between. A complete cycle consists of four pulses $XYXY$, thus the total time period of a cycle is $T_c=4\tau_d$. In the limit of $\tau_d \rightarrow 0$, the qubit can be stored for an arbitrarily long time.
The intuition behind DD-sequences is the average Hamiltonian theory. Let $U_C(t)=\mathcal{T} \exp \{-i\int_0^t dt' H_C(t')\}$ be the total unitary applied by $H_C(t')$ up to time $t$.
In the interaction picture defined by $U_C(t)$, the dynamics is governed by the Hamiltonian $\tilde{H}(t)=U_C^\dagger(t)H_0U_C(t)$.
If the time interval $\tau_d$ between pulses is much smaller than the time scale defined by the norm of $\| H_0\|$, it is reasonable to consider the average of $\tilde{H}(t)$ within a cycle.
The zeroth-order average Hamiltonian in $T_c$ (with respect to $\tau_d$) is
\begin{equation}
\bar{H}^{(0)}=\frac{1}{T_c}\int_0^{T_c} dt' U_C^\dagger(t)H_0U_C(t) \ .
\end{equation}
For the $XY_4$ sequences introduced above, it is easy to compute $\bar{H}^{(0)}=\frac{1}{4}\sum_{\sigma\in \{I,X,Y,Z\}} \sigma H_0 \sigma$.
Since the mapping $O\rightarrow \sum_{\sigma\in \{I,X,Y,Z\}} \sigma O \sigma$ maps any $2\times 2$ matrix to $0$, by linearity we know $\bar{H}^{(0)}=0$.

Here we are going to list several classes of DD-sequences. We will first explain how to concatenate two sequences, as most long DD-sequences are constructed in this manner.
Given two DD-sequences $A=P_1\cdots P_m$ and $B=Q_1\cdots Q_n$, the concatenated sequence $A[B]$ is
\begin{align*}
A [ B ]=(P_1Q_1)Q_2\cdots Q_n(P_2Q_1)Q_2\cdots Q_n \cdots (P_mQ_1)Q_2\cdots Q_n
\end{align*}
As an example, when we concatenate the length-2 and length-4 sequences $XX$ and $XYXY$, we obtain $IYXYIYXY$.

We will use $P_i$ to represent any Pauli matrix $X$, $Y$ or $Z$, and for $i\neq j$, $P_i \neq P_j$. The families of DD-sequences can then be listed as the following:
\begin{description}
	\small
	\item[DD4] length-4 sequences $P_1 P_2 P_1 P_2$.
	\item[DD8] length-8 sequences $I P_2 P_1 P_2 I P_2 P_1 P_2$.
	\item[EDD8] length-8 sequences $P_1 P_2 P_1 P_2 P_2 P_1 P_2 P_1$
	\item[CDD16] length-16 concatenated sequences $DD4[DD4]$
	\item[CDD32] length-32 concatenated sequences $DD4[DD8]$ and $DD8[DD4]$
	\item[CDD64] length-64 concatenated sequences $DD4[CDD16]$ and $DD8[DD8]$
\end{description}
Longer DD-sequences can again be obtained by the concatenation of the ones listed above, and in the ideal situation they provide better and better protection against the noise.
However, with realistic experimental capability, the performance usually saturates at a certain concatenation-level.
Since at this moment we are only optimizing short DD-sequences, the listed ones are sufficient to provide a baseline for our purpose.
One important family we did not include here is the ``Knill DD'' (KDD)~\cite{Ryan2010}, because it requires the use of non-Pauli gates.

However, we cannot expect these requirements to be met in all real world experiments. The two major imperfections that are often studied are the flip-angle errors and the finite duration of the pulses. Flip-angle errors arise from not being able to control the strength and time duration of $H_C(t)$ perfectly, thus the intended pulse $V(t)=O\delta (t)$ becomes $V(t)=(1\pm \epsilon) O\delta (t)$. And since zero-width pulses $O\delta (t)$ are experimentally impossible, we must consider finite-width pulses which approximate the ideal ones.
In this paper, we will only consider the imperfection of finite-width pulses.
However, it is straightforward to apply our algorithm to pulses with flip-angle errors.

\subsection{Measure of performance}
\label{sec_measure_of_performance}
There are multiple ways to quantify the performance of DD-sequences. In practice, we choose different measures to suit the intended applications.
Here we use the same measure as in~\cite{Quiroz2013}, which has the advantage of being (initially) state-independent and having a closed formula for numerical simulation:
\begin{align}
D(U,I)=\sqrt{1-\frac{1}{d_Sd_B}\| \trace_S (U)\|_{\trace} }
\end{align}
where $U$ represents the full evolution operator generated by $H(t)$, $d_S$ and $d_B$ are the dimensions of the system and environment Hilbert space $\mathcal{H}_S$ and $\mathcal{H}_B$, respectively.
$\| X \|_{\trace}=\trace (\sqrt{X^\dagger X})$ is the trace-norm, and $\trace_S (\cdot)$ is the partial trace over $\mathcal{H}_S$.
The smaller $D(U,I)$ is, the better the system preserved its quantum state after the time evolution.
For example, the ideal evolution $U=I_S\otimes U_B$ has the corresponding $D(U,I)=0$.

In experiments, it is very hard to evaluate $D(U,I)$, as we often do not have access to the bath's degree of freedom.
Instead, the performance of DD-sequences is often gauged by doing process tomography for the whole time duration where DD is applied~\cite{DeLange2010,Souza2012}.
Although it is a different measure compared to our choice above, the optimization procedure can still be applied as it does not rely on the concrete form of the measure.
Moreover, for solid state implementations such as superconducting qubits or quantum dots, a typical run of initialization, applying DD-sequences and measurements can be done on the time scale of 1 ms or much faster.
Thus, it is realistic that on the time scale of days we can gather a large dataset of DD-sequences and their performance, which is needed for our algorithm.

\subsection{Recurrent Neural Networks}

Sequential models are widely used in machine learning for problems with a natural sequential structure, e.g. speech and handwriting recognition, protein secondary structure prediction, etc..
For dynamical decoupling, not only do we apply the gates sequentially in the time domain, but also the longer DD sequences are often formed by repetition or concatenation of the short ones.
Moreover, once the quantum information of the system is completely mixed into the environment, it is hard to retrieve it again by DD.
Thus, an educated guess is that the performance of a DD-sequence largely depends on the short subsequences of it, which can be modelled well by the sequential models.

Since our goal is not simply to approximate the distribution of good dynamical decoupling sequences by learning their structure but to sample from the learned distribution to efficiently generate new good sequences, we will further restrict ourselves to the class of generative sequential models. Overall, these models try to solve the following problem: given $\{x_i\}_{i<t}$, approximate the conditional probability $p(x_t | x_{t-1},\ldots ,x_1 )$.
As a simple example, we can estimate the conditional probability $p(x_t | x_{t-1})$ from a certain data set, and use it to generate new sequences \footnote{This idea can be at least dated back to Shannon~\cite{shannon1948mathematical}, where this model generated ``English sentences'' like ``ON IE ANTSOUTINYS ARE T INCTORE ST BE S DEAMY ACHIN D ILONASIVE TUCOOWE AT....''}.
For more sophisticated problems (e.g. natural language or handwriting), it is not enough to only consider the nearest neighbour correlations as simple models like Markov-chains of order one do.

The long short-term memory (LSTM) network, a variation of the recurrent neural network (RNN), is a state-of-the-art technique for modelling longer correlations~\cite{Graves2013} and is comparably easy to train. The core idea of RNNs is that the network maintains an internal state in which it encodes information from previous time steps. This allows the model to, at least theoretically, incorporate all previous time steps into the output for a given time. Some RNNs have even been shown to be Turing-complete~\cite{pollack1987}. In practice, however, RNNs often can only model relatively short sequences correctly due to an inherently unstable optimization process. This is where LSTMs improve over normal RNNs, as they allow for training of much longer sequences in a stable manner. Furthermore, LSTMs, like all ANNs, are based on matrix multiplication and the element-wise application of simple non-linear functions. This makes them especially efficient to evaluate.

From the machine learning perspective, we treat the problem at hand as a supervised learning problem where we provide the model with examples that it is to reproduce according to some error measure.
It is also possible to formulate our problem in the framework of reinforcement learning. However, since we only compute the performance of a whole DD sequence, there is no immediate reward when choosing a gate in the middle of the sequence.
Given the length of the sequences we are optimizing, it is likely a reinforcement learning algorithm will need help from certain (un)supervised learning, similar to the way in~\cite{silver2016}.
A short introduction to machine learning, LSTMs and their terminology can be found in the appendix. More exhaustive discussions can be found in~\cite{Nielsen2015, greff2015, goodfellow2016}. 

\section{Algorithm}
\label{sec:algorithm}

\begin{algorithm}[t]
\setlength{\leftskip}{10pt}
\setlength{\skiprule}{10pt}
\caption{Optimization Algorithm}\label{em_algo}
    \SetKwInOut{Input}{Input}
    \SetKwInOut{Output}{Output}
    \SetKwFunction{RandData}{generateRandomData}
    \SetKwFunction{ModelData}{generateDataFromModels}
    \SetKwFunction{BestData}{keepBestData}
    \SetKwFunction{BestModels}{keepBestKModels}
    \SetKwFunction{TrainModels}{trainRandomModels}
    \SetKwFunction{TrainBestModels}{trainBestModels}

    \Input{Number of initial models to train: $n$, Number of models to keep: $k$, Percentage of data to keep: $p$, Set of possible topologies: $\mathcal{M}$, Size of data: $d$}
    $D \leftarrow$ \RandData{d} \;
    $D, \langle \varsigma_s \rangle \leftarrow$ \BestData{D,p} \;
    $M \leftarrow$ \TrainModels{n, D, $\mathcal{M}$} \;
    $M \leftarrow$ \BestModels{M,k} \;
    \While{$\langle \varsigma_s \rangle$ not converged}{
    $M \leftarrow$ \TrainBestModels{D} \;
    $D \leftarrow$ \ModelData{M, d} \;
    $D, \langle \varsigma_s \rangle \leftarrow$ \BestData{D,p} \;
    }
    
  	\Output{$\langle \varsigma_s \rangle, D, M$}
\end{algorithm}

The algorithm presented in this section is designed with the goal in mind to encode little prior knowledge about the problem into it, in order to make it generally applicable to different imperfections in the experiment. Following this idea, the method is agnostic towards the nature of the considered gates, the noise model and the measure of performance. To implement this, the algorithm assumes that
\begin{itemize}
\item the individual gates are represented by a unique integer number such that every sequence $s \in  \mathcal{G}^{\otimes L_s}$ with $\mathcal{G}$ denoting the set of the unique identifiers and $L_s$ being the length of $s$.
\item it is provided with a function $f(s)$ to compute the score $\varsigma_s$ of a given sequence $s$, taking into account the noise model.
\end{itemize}
The optimization problem we want to solve is 
\begin{align}
\min_s f(s) = \min_s  \varsigma_s.
\end{align}
By assumption, we have no information about $f$ but can efficiently evaluate it. We do furthermore assume the set of good sequences to exhibit common structural properties that can be learned well by a machine learning model.
So, we propose to solve it indirectly by training a generative model $m \in \mathcal{M}$ to approximate the distribution of good sequences, $\mathcal{M}$ being the set of possible models. That means we assume $s_t \sim p_m(s_{t-1},\dotsc,s_1)$ with $s_t$ being the gate at time $t$ and $p_m$ denoting the distribution learned by $m$. Then, we want to find an optimal $m$ that ideally learns a meaningful representation of the structure of good sequences. In this work we choose the type of model to be the LSTM.
We now tackle this surrogate problem by alternatingly solving
\begin{align}
\max_{m \in \mathcal{M}} \mathcal{L}(m|T), 
\end{align}
where $\mathcal{L}$ denotes the likelihood and $T$ the training data, and then sampling sequences from the model $m$ to generate a new $T$ consisting of better solutions. The algorithm hence consists of two nested optimization loops, where the inner loop fits a number of LSTMs to the current data while the outer loop uses the output of the inner loop to generate new training data. This scheme of alternatingly fixing the data to optimize the models and consecutively fixing the models to optimize the data resembles the probabilistic model building genetic algorithm~\cite{pelikan2002survey} and to some extent the expectation-maximization algorithm~\cite{dempster1977}.
The method is shown in Algorithm~\ref{em_algo}. Partial justification of this heuristic algorithm is given in the appendix~\ref{sec_compare_opti_algo}.
However, it is easy to see that the algorithm will not always find the global optimum.
For example, it is conceivable that for certain problems the second to the 100-th best solutions share no common structure with the first one. In that case, it would be unlikely for the machine learning approach to find the optimal one.
There is however likely no universal method to bypass this obstruction, as unless we know the best sequences already, it is impossible to verify that they exhibit some structure similar to the training sets.
This obstruction seems natural since many optimization problems are believed to be computationally hard. Thus, we should not assume be able to solve them by the above routine.

We will now explain the most important aspects of the algorithm in more detail:
\paragraph{Choice of LSTMs} The data we want to generate in our application is of sequential nature. This makes employing LSTMs an obvious choice as they pose one of the most powerful models available today for sequential data. Furthermore, the known well-performing families of DD sequences are constructed by nested concatenations of shorter sequences and hence show strong local correlations as well as global structure. LSTMs and especially models consisting of multiple layers of LSTMs are known to perform very well on such data and should therefore be able to learn and reproduce this multi-scale structure better than simpler and shallow models.

\paragraph{Generation of the initial training data} The size $d$ and the quality, i.e. the percentage $p$ of the initial data to be kept, are the parameters which we can specify. The data are then generated by sampling a gate from the uniform distribution over all gates for each time-step. The average score of the initial data can then be used as a baseline to compare against in case no other reference value is available.
We would like to point out that in the application considered in this work, an alternative way to generate the initial data might be to use the models trained on shorter sequences. This approach could lead to an initial data set with much higher average score, but at the price of introducing the bias from the previously trained RNNs.

\paragraph{Training of the LSTMs} To reduce the chance of ending up in a bad local optimum, for each training set several different architectures of LSTMs are trained (see~\ref{RNN} for detailed description of LSTMs). These models are independently sampled $\mathcal{M}$. More precisely, for the first generation of models, we sample a larger set of $n$ models from $\mathcal{M}$ and train them. We then select the best $k$ models and reuse them for all following generations. While it might introduce some bias to the optimization, this measure drastically reduces the number of models that need to be trained in total.
The training problem is defined by assuming a multinoulli distribution over the gates of each time step and minimizing the corresponding negative log-likelihood $-\sum_t \delta_{s_t,i} \log p_{m,i}(s_{t-1},\dotsc,s_1)$, where $i$ is the index of the correct next gate, $p_{m,i}$ is its predicted probability computed by the LSTM $m$ and $\delta_{s_t,i}=1$ iff $s_t=i$. This error measure is also known as the \emph{cross-entropy}. To avoid overfitting, we use a version of early stopping where we monitor the average score $\langle \varsigma_s \rangle_{p_m}$ of sequences generated by $m$ and stop training when $\langle \varsigma_s \rangle_{p_m}$ stops improving. We employ the optimizer Adam~\cite{kingma2015} for robust stochastic optimization.

\paragraph{Selecting the best models} As we employ early stop based on the average score $\langle \varsigma_s \rangle_{p_m}$, we also rank every trained model $m$ according to this measure. One could argue that ranking the models with respect to their best scores would be a more natural choice. This however might favour models that actually produce bad sequences but have generated a few good sequences only by chance. Using $\langle \varsigma_s \rangle_{p_m}$ is hence a more robust criterion. It would of course be possible to also consider other modes of the $p_m$, like the variance or the skewness. These properties could be used to assess the ability of a model to generate diverse and good sequences.
We find however that the models in our experiments are able to generate new and diverse sequences, thus we only use the average score as benchmark for selecting models.

\paragraph{Generation of the new training data} The selected models are used to generate $d$ new training data by sampling from $p_m$. This is done by sampling $s_t$ from $p_i(s_{t-1},\dotsc,s_1)$ beginning with a random initialization for $t=1$ and then using $s_{t-1}$ as input for time step $t$. We combine the generated sequences with the previous training sets, remove any duplicates, and order the sequences by their scores. We then choose the best $p$ percent for the next iteration of the optimization. This procedure ensures a monotonic improvement of the training data. Note that all selected models contribute equally many data to strengthen the diversity of the new training data. A possible extension would be to apply weighting of the models according to some properties of their learned distributions. Note though that ordering the generated sequences by their score is already a form of implicit weighting of the models.





\section{Numerical Results}
\label{sec_numerical_results}
\subsection{Noise model and the control Hamiltonian}
\label{sec:noise_and_control_Ham}
Throughout the paper, we will use the same noise model as in~\cite{Quiroz2013}. We consider a 1-qubit system and a 4-qubit bath, namely $\textsf{dim} (\mathcal{H_S})=2$ and $\textsf{dim} (\mathcal{H_B})=16$. The small dimension of the bath is for faster numerical simulation, and there is no reason for us to think that our algorithm would only work for a small bath as the size of the bath enters the algorithm only via the score-computation function.
The total noise Hamiltonian consists of (at most) 3-body interactions between the system and bath-qubits with random strength:
\begin{align}
\label{eq_total_noise_Hamiltonian}
H_0=\sum_{\mu \in \{I,X,Y,Z\}} \sigma^\mu \otimes B_{\mu} \ ,
\end{align}
where $\sigma^\mu$ is summed over Pauli-matrices on the system-qubit. And $B_{\mu}$ is given by
\begin{align*}
B_\mu= \sum_{i\neq j}\sum_{\alpha,\beta} c^\mu_{\alpha\beta} \left(\sigma_i^\alpha \otimes \sigma_j^\beta \right) \ ,
\end{align*}
where $i,j$ is summed over indices of the bath qubits, and $\sigma_i^{\alpha (\beta)}$ is the Pauli-matrix on qubit $i$ of the bath.
We consider the scenario where the system-bath interaction is much stronger than the pure bath terms.
More precisely, we set $c^\mu_{\alpha\beta}\approx 1000 c^I_{\alpha\beta}$ for $\mu\in \{X,Y,Z\}$.
Apart from this constraint, the absolute values $| c^\mu_{\alpha\beta} |$ are chosen randomly from a range $[a,b]$, where we set $b\approx 3a$ to avoid too many terms vanishing in ~\eqref{eq_total_noise_Hamiltonian}.
The result Hamiltonian has a 2-norm $\|H_0\|=20.4$.

For the control Hamiltonian, we consider the less explored scenario where the pulse shape have finite width but no switch time between them (100\% duty cycle). In other words, the control Hamiltonian is piecewise constant
\begin{align*}
H_C (t)= H_k\ , \quad \text{for}\,\, k\tau_d \leq t <(k+1)\tau_d \ ,
\end{align*}
where $\tau_d$ is a small time period with respect to the norm of $H_0$, and $e^{-iH_k\tau_d}\in \{I,X,Y,Z\}$. This is a good toy model for experimental settings whose DD-performance is mainly limited by the strength of the control Hamiltonian, but not the speed of shifting between Hamiltonians.
Since this regime is less explored in theoretical studies, it is an interesting scenario to explore via machine learning.
Another restriction we put on $H_C (t)$ is
\begin{align}
H_C (t)= - H_C (T-t) \ ,
\end{align}
where $T$ is the total evolution time.
This condition ensures $U_C(T)=\mathcal{T} \exp \{-i\int_0^T dt' H_C(t')\}=I$, and it allows us to apply the same code on the setting where the system has more than one qubit.
It is known that this family of symmetric Hamiltonians can remove the first order terms of $\tau_d$ in the average Hamiltonian\cite{Viola1999,Souza2012a}.
So strictly speaking, this should be counted as prior knowledge.
However, when we compare the known DD-sequences with the numerically found ones, we also use the symmetric version of the known DD-sequences.
Thus, we perform the comparison on equal terms.

In the following, we present the results of a number of experiments we have conducted to evaluate the performance of our method. We consider sequences consisting of 32, 64 and 128 gates for varying values of $\tau_d$. This translates to having to optimize the distribution of the first 16, 32 and 64 gates respectively. To compute $\varsigma_s$, we use the figure of merit $D$ as defined in Section~\ref{sec_dynamical_decoupling}. Thus, a lower score is better.
For $\mathcal{M}$, we consider models with two or three stacked LSTM-layers followed by a final softmax layer. The layers comprise 20 to 200 units where layers closer to the input have a higher number of units. We allow for peephole connections and linear projections of the output of every LSTM-layer to a lower number dimensions~\cite{greff2015}. 
The optimization parameters are also randomly sampled from sets of reasonable values. We choose the step rate to be in $\{10^{-1},10^{-2}\}$ and the batch size to take values in $\{200, 500, 1000\}$. The parameters specific to the Adam optimizer $\beta_1, \beta_2$ and $\epsilon$, we sample from $\{0.2, 0.7, 0.9\}$, $\{0.9, 0.99, 0.999\}$ and $\{10^{-8}, 10^{-5}\}$ respectively. We perform a truncation of the gradients to 32 time steps in order to counter instabilities in the optimization (see~\ref{opti_rnn}). As we have stated above, we also employ early stopping in the sense that, for every optimization of a model, we keep the parameters that generate the sequences with the best average score. The algorithm was run until either the best known score was beat or the scores converged, depending on the goal of the respective experiment. 
We will now briefly list the concrete experiment settings and discuss the results.
\renewcommand{\arraystretch}{1.3}
\begin{table}
	\caption{A comparison of the results obtained in experiments E1, E2, E3 and E4 to the best theoretically derived DD families. For each experiment, the average and best score of the last training data and the average score of the best model of the last generation are shown. They are compared to random sequences and the two DD classes that yield the best average and overall best score respectively. 
		The best results are printed bold.}
	\centering
	\begin{minipage}{.47\textwidth}
		\subcaption{Experiment E2}
		\begin{tabular}{c *{2}{c} }
			\toprule
			Sequences & $\langle \varsigma_s \rangle$ & $\min \varsigma_s$ \\ \midrule
			EDD8 & 0.002398 &  0.002112 \\ 
			CDD32 & 0.053250 &  0.000803 \\ 
			Last training set E2 & \textbf{0.000712} &  \textbf{0.000381} \\
			Best model E2 & 0.016692 & -  \\ 
			Random & 0.341667 &  - \\ 
			\bottomrule
		\end{tabular}
		\label{result_tables_E2}
	\end{minipage}
	\hfill
	\begin{minipage}{.48\textwidth}
		\subcaption{Experiment E3}
		\begin{tabular}{c *{2}{c} }
			\toprule
			Sequences & $\langle \varsigma_s \rangle$ & $\min \varsigma_s$ \\ \midrule
			EDD8 & 0.004793 &  0.004222 \\ 
			CDD64 & 0.031547 &  0.001514 \\
			Last training set E3 & \textbf{0.000827} &  \textbf{0.000798} \\ 
			Best model E3 & 0.029341 & -  \\ 
			Random & 0.44918 &  - \\ 
			\bottomrule
		\end{tabular}
		\label{result_tables_E3}	
	\end{minipage}
	\hfill
	\subcaption{Experiments E1 and E4}
	\begin{tabular}{c *{2}{c}}
		\toprule
		Sequences & $\langle \varsigma_s \rangle$ & $\min \varsigma_s$ \\ \midrule
		EDD8 & 0.000151 & 0.000133 \\	
		CDD16 & 0.010699 &  0.000074 \\
		Last training set E1 & \textbf{0.000112} &  \textbf{0.000070} \\
		Last training set E4 & 0.007178 &  0.000082 \\
		Best model E1 & 0.003089 & -  \\
		Random & 0.125371 &  - \\ 
		\bottomrule
	\end{tabular}
	\label{result_tables_E1}
	
\end{table}

\begin{figure}
	\begin{subfigure}[t]{0.5\textwidth}
		\centering\captionsetup{width=.8\linewidth}
		\includegraphics[width=1\textwidth]{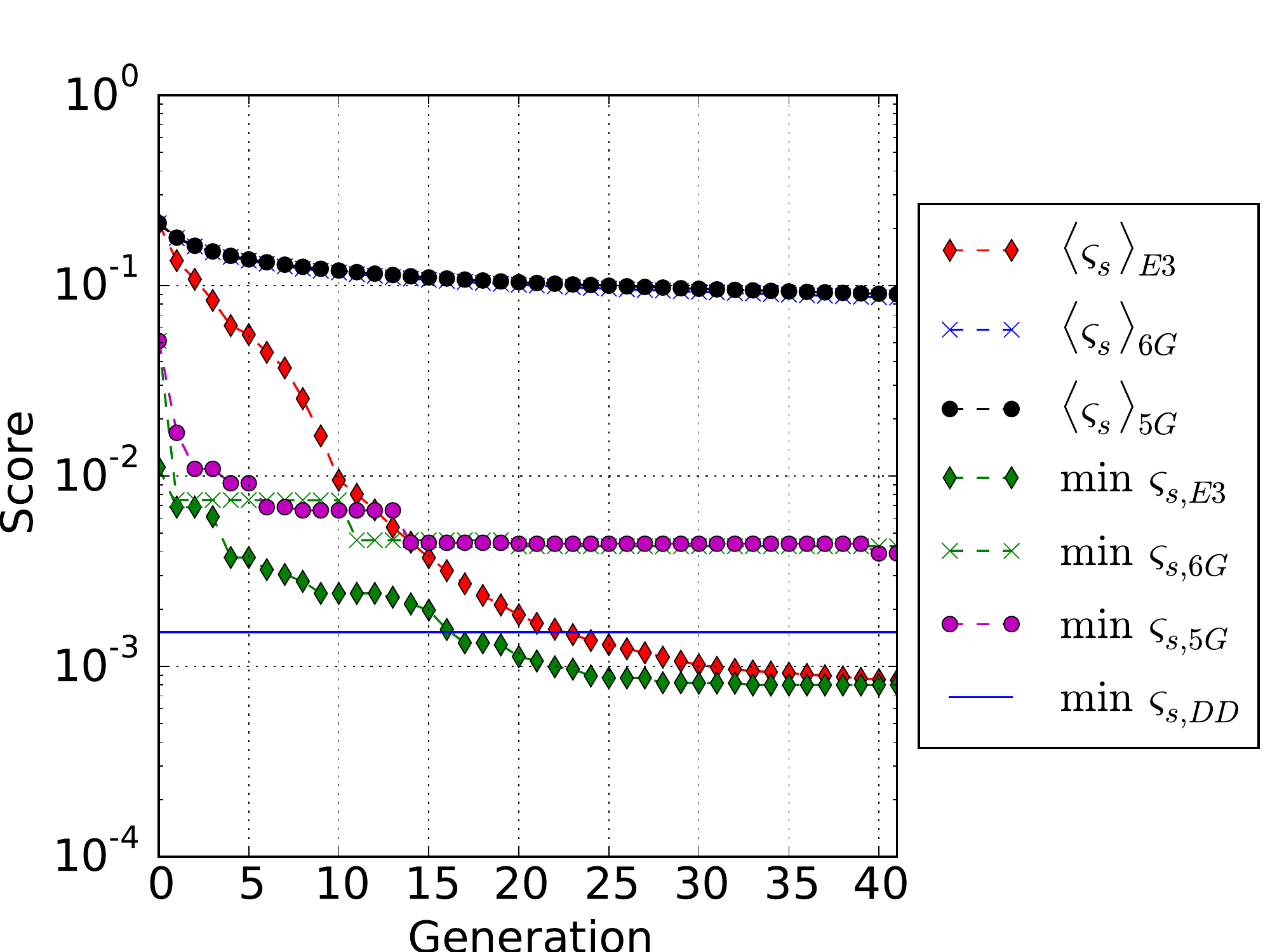}
		\subcaption{Experiment E3}
		\label{result_plots_E3}
	\end{subfigure}
	\begin{subfigure}[t]{0.5\textwidth}
		\centering\captionsetup{width=.8\linewidth}
		\includegraphics[width=1\textwidth]{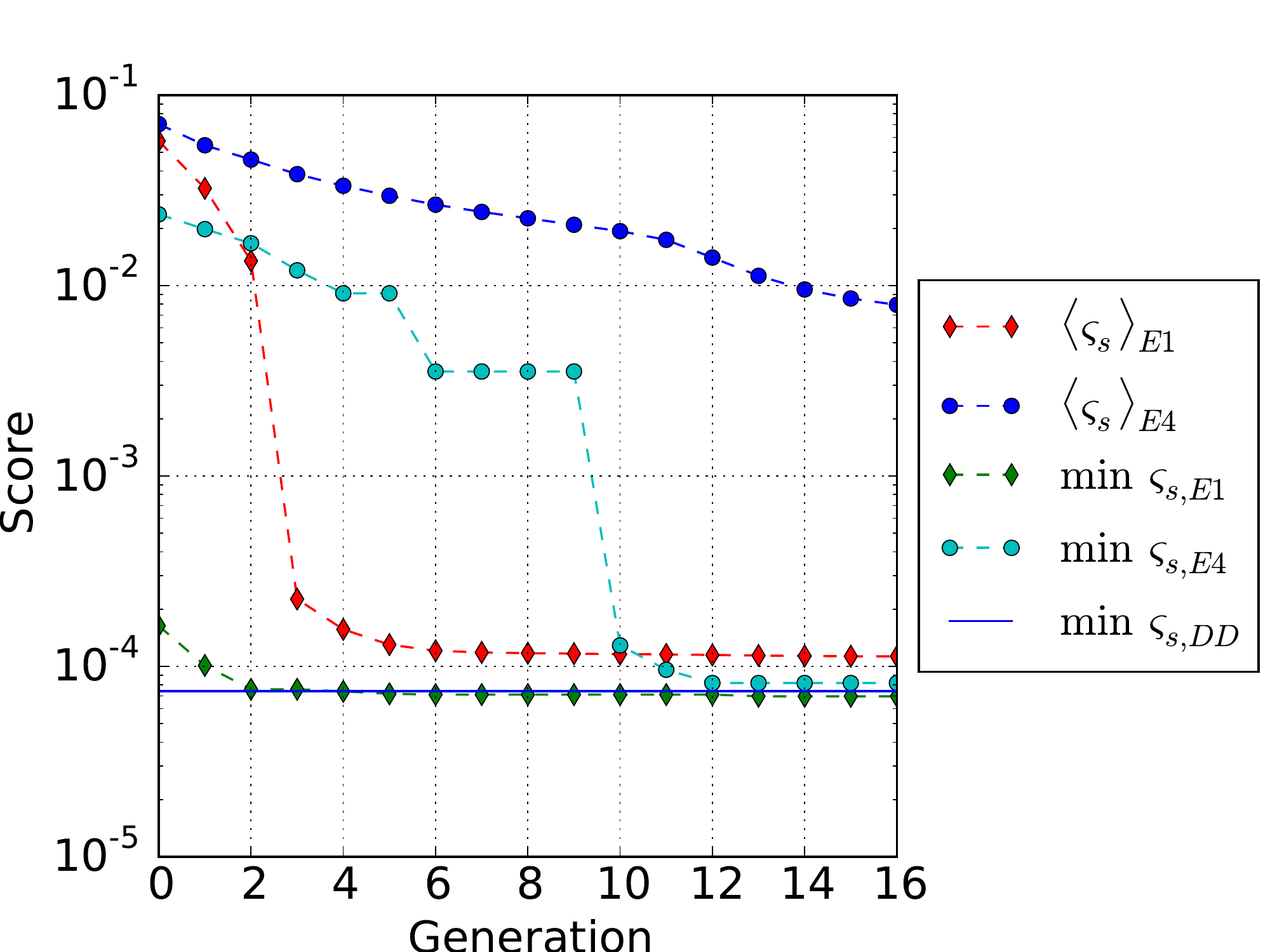}
		\subcaption{Experiment E1 and E4}
		\label{result_plots_E1}
	\end{subfigure}

	\caption{Two figures showing the convergence of the algorithm a) in E3 compared to the case where LSTMs are replaced by 5/6-gram models and b) in E1 comparted to E4 as both consider the same problem setting. In a) it is clearly visible that LSTMs outperform the n-gram models while b) reflects the physical knowledge that the Pauli unitaries are a better choice than random gates. As a reference, we show the score of the best DD sequence obtained from the known DD classes.}
	\label{result_plots}
\end{figure}

\paragraph{Exp. E1: Length 32}
In this first experiment, we considered sequences of 32 gates with $\tau_d= 0.002$. We let the algorithm train $n=30$ models initially and set the number of models to be kept $k$ to 5. We combined the data generated by the LSTMs with the previous training set after each generation, and chose the best 10\% as the new training data, consisting of 10,000 sequences for each generation. We let every model train for 100 epochs.

\paragraph{Exp. E2: Length 64} In our next experiment, we tackled a more difficult scenario with 64 gates and a larger $\tau_d=0.004$. We set $n=50$ and $k=5$. Again, we used the best 10\% of both generated and previous data as new training data which consists in total 10,000 sequences for each training set. 

\paragraph{Exp. E3: Length 128} In the third experiment we tried our method on even longer sequences of 128 gates with $\tau_d$ again being 0.004. Due to the very large sequence space, we set the size of the training sets to 20,000, again using the best 10\% of sequences generated by the selected models and the previous training set. The number of epochs was increased to 200.
We set $n=30$ and $k=5$. Here, we let the algorithm run until both average and best score converged to examine its behaviour in long runs.

\paragraph{Exp. E4: Length 32 with Random Gates} Finally, we tested the performance of Algorithm~\ref{em_algo} in the case where we replaced the Pauli gates $\{I,X,Y,Z\}$ with ten randomly chosen gates. More precisely, we chose each gate $g_j$ to be a randomly generated single two-dimensional unitary operator with eigenvalues $1$ and $-1$, i.e.
$g_j=U_j^{\dagger} X U_j$, where $U_j$ is a random unitary. All other parameters were kept as in experiment E1.

In the Tables~\ref{result_tables_E3}, ~\ref{result_tables_E2} and ~\ref{result_tables_E1}, we compare the last training data and the best model of the last generation of E1-E4 against the two DD families that achieve the best average and minimal scores for the given experiment respectively. We also plot the convergence of the training data of E3 and E1 with E4 in the Figures~\ref{result_plots_E3} and \ref{result_plots_E1} respectively.
In general, the results for E1, E2 and E3 clearly show that our method outperforms DD, achieving a better minimal score of the generated data in a moderate number of iterations and with a relatively small set of models. The results of E4 will be discussed below.
These findings indicate that our method converges to good local optima and that the models are able to learn a meaningful internal representation of the sequences that allows for efficient sampling of good sequences. There is however a noticeable gap between the scores of the training data and the models. A possible remedy for this could be an increase of the training data size or an adjustment of the model parameters in later stages of the optimization to account for the change in the structure of the data.

To assess the importance of LSTMs for the performance of our algorithm, in experiment E3, we also ran a different version of our method where we replaced the LSTMs by simple 5/6-gram models, which only model and generate sequences based on local correlations (see Appendix~\ref{sec:ngram} for the definition). The convergence plots in Figure~\ref{result_plots_E3} show that LSTMs are indeed superior to the simpler models. They are able to improve the average and best scores faster and ultimately let the algorithm converge to a better local optimum. This advantage most likely results from the fact that the LSTM-models are able to leverage information about longer range correlations in the data. These results hence justify our choice of LSTMs as machine learning model to optimize DD-sequences.

We also compared the results of experiments E1 and E4 to examine the importance of using the Pauli group as the gate set.  Figure~\ref{result_plots_E1} shows that while for E1 the average score quickly becomes very good and the best score exceeds the best known result after a few generations, in E4 the average score of the data improves much slower and remains significantly worst than that of E1. Although the best score exhibits a much stronger improvement, it eventually converges to a value slightly worse than that of the best theoretical DD-sequence and the one found in E1. This is expected since with the Pauli group we can achieve first-order decoupling with DD sequences of length 4, which is the shortest. On the other hand, with random unitaries, in general it will take much longer sequences to have approximate first-order decoupling, during which the system and environment can become fairly entangled.

Another interesting aspect to note is the rather strong improvement of the average scores occurring in E3 and E1 between generations 8 to 10 and 2 to 3, respectively. These jumps can be explained by the known existence of several strictly separate regimes in sequence space that differ strongly in their performance. The results indicate that our algorithm is able to iteratively improve the learned distributions to eventually capture the regime of very good sequences.

In order to verify that sampling the initial training data from the distributions learned for shorter sequences is a viable alternative to uniform sampling, we let the best model obtained in E2 generate an initial data set for the problem setting of E3. The obtained data was found to have an average score of 0.037175, which is about one order of magnitude better than the average of the initial training data generated by uniform sampling.
\section{Conclusion}

We have introduced a novel method for optimizing dynamical decoupling sequences, which differs from previous work by the ability to utilize much larger datasets generated during the optimization. Its ability to efficiently generate large sets of good sequences could be used along with other optimization methods to cover their weaknesses or to perform statistical analysis of these sequences.
We showed that for certain imperfect control Hamiltonians, our method is able to outperform (almost all) known DD-sequences.
The little prior knowledge about DD we use is (1) choosing Pauli operators as pulses in the sequences (see experiment E4 and its discussion), (2) choosing specific lengths for the DD-sequences and (3) enforcing the reversal symmetry, as discussed in section~\ref{sec:noise_and_control_Ham}.
However, we do not need to initialize the dataset in a specific way as in the Appendix C.5.a of~\cite{Quiroz2013}, which actually contains a certain amount of prior knowledge of DD. Also, our method does not fundamentally rely on the prior knowledge stated above.
It is conceivable that the use of this prior knowledge can be lifted, at the price of a possibly much slower optimization procedure.
For example, the KDD scheme helps to further increase the performance of CDD-sequences in some experiments~\cite{Souza2011}.
Thus, an interesting question is when given the freedom of applying non-Pauli gates and choosing variable lengths of the sequences, whether our algorithm could discover a similar strategy.
Thus, a possible direction of future research is to see how we can minimize the slow-down when not incorporating any prior knowledge and whether we can obtain good DD-sequences with non-Pauli pulses.

While we have applied the algorithm to the case of quantum memory and compared it to dynamical decoupling, it is of general nature. It can in principle be applied to every problem where the optimization of a sequence of gates with respect to some well-defined figure of merit is desired and where it is feasible to evaluate this performance measure for larger numbers of sequences. However, due to the nature of the underlying machine learning model, good results will likely only be obtained for problems whose solution depends strongly on local correlations in the sequences.


\section*{Acknowledgements}

We would like to thank Geza Giedke for helpful discussions and comments on the draft. The idea of this paper partially stems from a discussion between Courtney Brell and Xiaotong Ni about using genetic algorithms to optimize quantum memory. We would also like to thank Peter Wittek for helpful comments.


\bibliography{ref}
\clearpage

\appendix



\section{Analysis}
\subsection{Local correlations of DD sequences}
As we suggested earlier, the reason we use RNNs as the probabilistic model is that the performance of dynamical decoupling sequences heavily depends on their local correlations. To illustrate this fact, we can count the frequency of length-2 (3) subsequences from the training set of the 30th generation in Experiment 3. 
We can then compare these statistics to the ones of the sequences generated by the LSTM, which is trained based on the training set.
We can see indeed the percentages match very well. To get more detail about local correlations, we could also count the frequency of length-3 subsequences (see table~\ref{tab:3-subsequences}).
Note that since the table is based on the datasets in the late stage of the optimization, the distribution of the subsequences are already very polarized. However, we observe the same behavior (the percentages matches well) in other experiments at different stages of the optimization as well.
\begin{table*}
\begin{ruledtabular}

		\begin{tabular}{c *{4}{|c} |}
			\diagbox[width=8em]{Previous}{Next gate} & I & X & Y & Z \\ \hline
			I & 0.00\% (0.00\%) & 0.04\% (0.08\%) & 0.15\% (0.68\%) & 0.02\% (0.08\%) \\ \hline
			X & 0.05\% (0.22\%) & 5.38\% (5.04\%) & 30.53\% (30.47\%) & 1.39\% (1.26\%) \\ \hline
			Y & 0.07\% (0.20\%) & 30.17\% (30.47\%) & 18.40\% (18.61\%) & 5.84\% (5.50\%) \\ \hline
			Z & 0.01\% (0.02\%) & 1.90\% (1.68\%) & 5.75\% (5.42\%) & 0.30\% (0.27\%) 
			
		\end{tabular}
		\caption{The frequency of length-2 subsequences, from the training set and the set generated by the trained LSTM (given in parentheses) at the generation 30 of Experiment 3. The total number of subsequences is around 1.2 million}
		\label{tab:2-subsequences}
\end{ruledtabular}
\end{table*}

\begin{table*}
	\begin{ruledtabular}
		\begin{tabular}{c *{4}{|c} |}
			\diagbox[width=8em]{Second}{Last gate} & I & X & Y & Z \\ \hline
			I & 0.00\% (0.00\%) & 0.02\% (0.05\%) & 0.12\% (0.55\%) & 0.00\% (0.01\%) \\ \hline
			X & 0.00\% (0.00\%) & 1.40\% (1.22\%) & 11.99\% (11.52\%) & 0.32\% (0.32\%) \\ \hline
			Y & 0.15\% (0.47\%) & 44.79\% (45.09\%) & 33.39\% (33.54\%) & 4.11\% (3.85\%) \\ \hline
			Z & 0.01\% (0.01\%) & 2.38\% (2.14\%) & 1.05\% (0.98\%) & 0.28\% (0.26\%) 

		\end{tabular}
		\caption{The frequency of length-3 subsequences started with the gate $X$, from the training set and the set generated by the trained LSTM (given in parentheses) at the generation 30 of Experiment 3. The total numbers of the subsequences started with $X$ are around 450 thousands.}
		\label{tab:3-subsequences}
\end{ruledtabular}
\end{table*}

However, RNNs do not only take into account local correlations, as we show in Figure~\ref{result_plots} that they perform better compared to the $n$-gram models, which we will introduce in the next subsection.

\subsection{$n$-gram models}
\label{sec:ngram}

$n$-grams are the simplest sequential models that treat the sequences as stationary Markov chains with order $n-1$. Operationally, given a set of sequences, we first estimate the conditional probability distribution 
\begin{align*}
p_{x_n, x_{n-1} \cdots x_1}= \mathsf{Pr}(X_t=x_n | X_{t-1}=x_{n-1}, \ldots X_{t-n+1}=x_1 ).
\end{align*}
Note that we assume the conditional probability is independent of $t$ (hence stationary Markov chain). The estimation is done by counting over the whole set of sequences.
The generation of new sequences based on the conditional probability $p_{x_n, x_{n-1} \cdots x_1}$ is straightforward, as we can repeatedly sample from it based on the previous $n-1$ items.
This behavior is different compared to the RNNs', which have memory units that can store information for arbitrary long time in theory.

\subsection{Optimization without reusing data from previous training sets}
\begin{figure}
	\includegraphics[width=.5\textwidth]{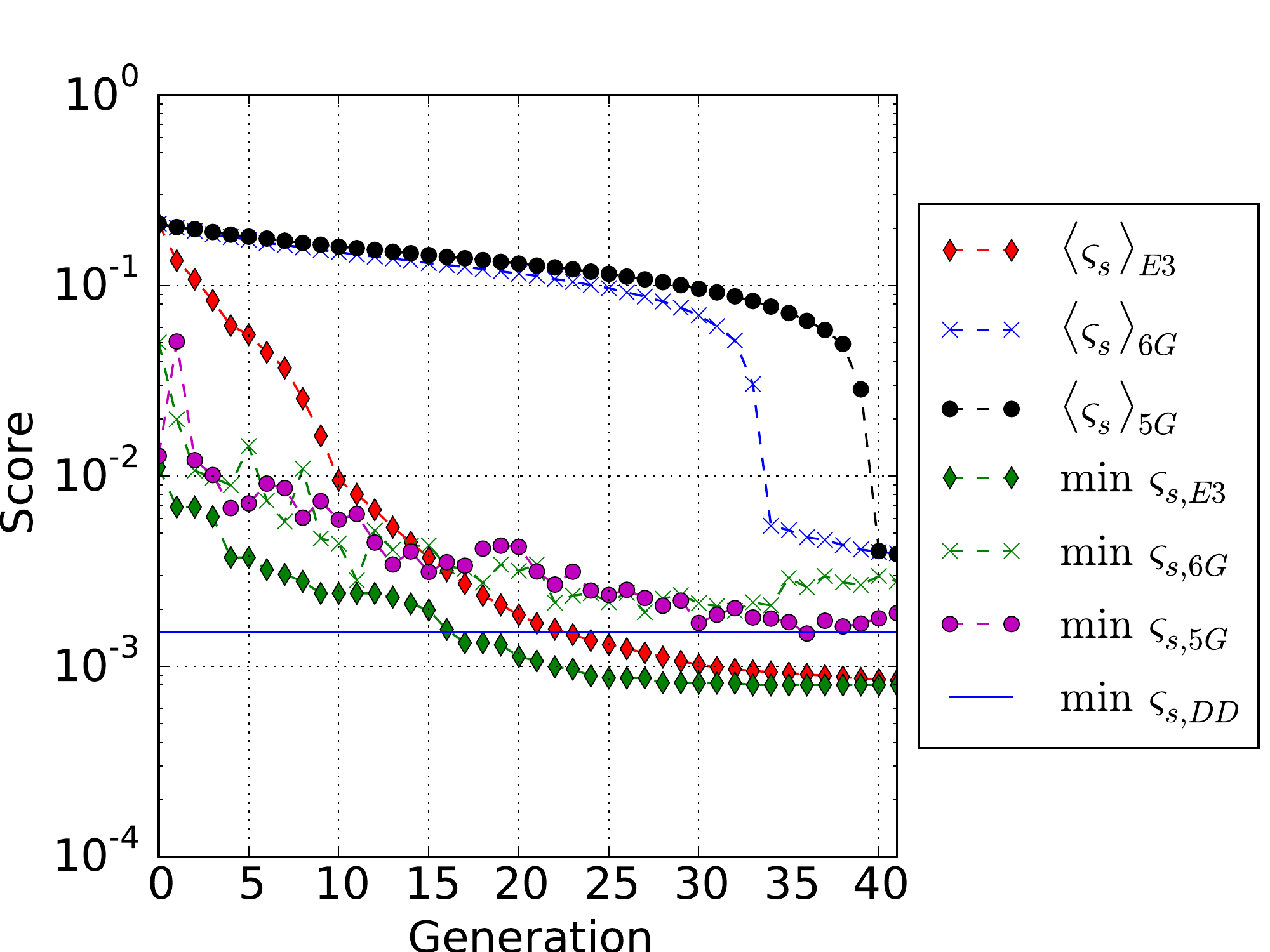}
	\caption{Experiment 3 and 5/6-gram without data reusage. Otherwise, the experiments are done in the same way as in Figure~\ref{result_plots_E3}.}
	\label{fig_reuse_data}
\end{figure}

During the optimization processes in the main text, we always reuse the data from previous training sets, in the sense that we first add the new sequences generated by the models to the training sets and then delete the worst sequences.
An interesting question is what will happen if we generate new training sets completely from the trained models.
In Figure~\ref{fig_reuse_data}, we plot the counterpart of Figure~\ref{result_plots_E3} with this modification (as well as not deleting duplicated sequences from the training set).
We can see that for the LSTMs experiment, the final minimum score gets slightly worse, which is 0.000874.
However, the 5/6-gram experiments actually performs better when not reusing data.
While it seems counterintuitive, this can be possibly explained by the fact that in the case of reused data with unique sequences the higher diversity of the data might make it harder for the models to find local correlations which then in turn slows down the optimization.  
There is other interesting information contained in the plot. For example, we can see the minimum scores almost always decrease, which implies that the LSTMs are able to learn new information about good sequences in most generations.

\subsection{Performance of the obtained sequences with a larger heat bath}
\label{sec:larger_bath}
In the main text, all the numerical simulations are done on a randomly generated noise Hamiltonian with the dimension of the bath being $\textsf{dim} (\mathcal{H_B})=16$.
The small dimension of the bath is used in order to have a fast simulation.
Here, we test the performance of some obtained sequences from the experiment 2, in the presence of a larger bath with $\textsf{dim} (\mathcal{H_B})=128$.
Apart from the change of dimension, the Hamiltonian $H_0$ is again randomly generated according to the description in~\ref{sec:noise_and_control_Ham}, which has a 2-norm $\|H_0\|=24.0$.
We then computed the scores of the top 500 DD sequences in the last generation of Experiment 2. The results are shown in Table \ref{tab_exp_h8}.
While the best score of the obtained sequences is worse than best score of CDD32, it is clear that on average, the obtained sequences still work fairly well.
This also suggests that our algorithm is potentially capable of adapting to the particular noise Hamiltonian, as the learned sequences outperform known DD-families in Experiment 2.
\begin{table}
	\begin{center}
		\begin{tabular}{c *{2}{|c} |}
			Sequences & $\langle \varsigma \rangle$ & $\min \varsigma$ \\ \hline
			EDD8 & 0.002781 &  0.002203 \\ \hline
			CDD32 & 0.053753 &  0.000432 \\ \hline
			Top 500 sequences & 0.001081 &  0.000626 \\ \hline
		\end{tabular}	
	\end{center}
	\caption{A comparison between the scores of the top 500 DD sequences in the last generation of Experiment 2 and some DD families for the larger bath $\textsf{dim} (\mathcal{H_B})=128$. The best score of the 500 sequences is worse than best score of CDD32.
	However, it is clear that on average, the obtained sequences still work fairly well.}
	\label{tab_exp_h8}
\end{table}

\section{Best Sequences}
We list here the best sequences we found in Experiment 1,2 and 3 from the numerical results section. We denote the identity by $I$, $X,Y,Z$ refer to the respective Pauli-matrices. Note that we show only the first half of the complete sequence as the second one is just the first half reversed.

\noindent \textbf{Experiment 1} X, Y, X, Z, X, Y, X, Z, Z, X, Y, X, Z, X, Y, X

\noindent \textbf{Experiment 2} Z, Z, X, Z, Z, Z, X, Z, Z, X, Z, X, X, X, Z, X, X, X, Z, X, X, Z, X, X, X, Z, X, Z, Z, X, Z, Z

\noindent \textbf{Experiment 3} Z, X, Z, Z, Y, X, Y, Z, Y, X, Y, X, Y, Y, X, Y, Y, Y, Y, X, Y, Y, Y, X, Y, Y, X, Y, X, Y, X, Y, Y, Z, X, Z, Y, Z, X, Z, Y, X, Y, X, X, Y, X, Y, X, Y, X, Y, Y, X, Y, Y, Y, X, Y, X, X, Y, X, X

\section{Comparison of optimization algorithms}
\label{sec_compare_opti_algo}

In this section, we will give a comparison between several optimization algorithms applied to black-box problems. In other words, the algorithm needs to optimize (minimize) the objective function $f$ only by looking at the values of $f(x)$ (without knowing the concrete formula of it). We are going to look at the following types of algorithms:
\begin{itemize}
	\item Gradient-based algorithms (when we can access the gradient of $f$), e.g. Newton's method, variants of gradient descent.
	\item Metropolis-Hasting algorithms and its variants, e.g. simulated annealing
	\item Genetic algorithm and its variants, e.g. probabilistic model building genetic algorithm (PMBGA).
\end{itemize}
The performance of an optimization algorithm depends heavily on the class of the problems it is applied to. (This fact is remotely related to the ``no free lunch theorem for optimization''). Thus in the following, we will use different objective functions to illustrate the strong and weak points of those algorithms.

\subsection{Gradient based algorithms}
\begin{figure}
	\centering
	\includegraphics[width=0.6\linewidth]{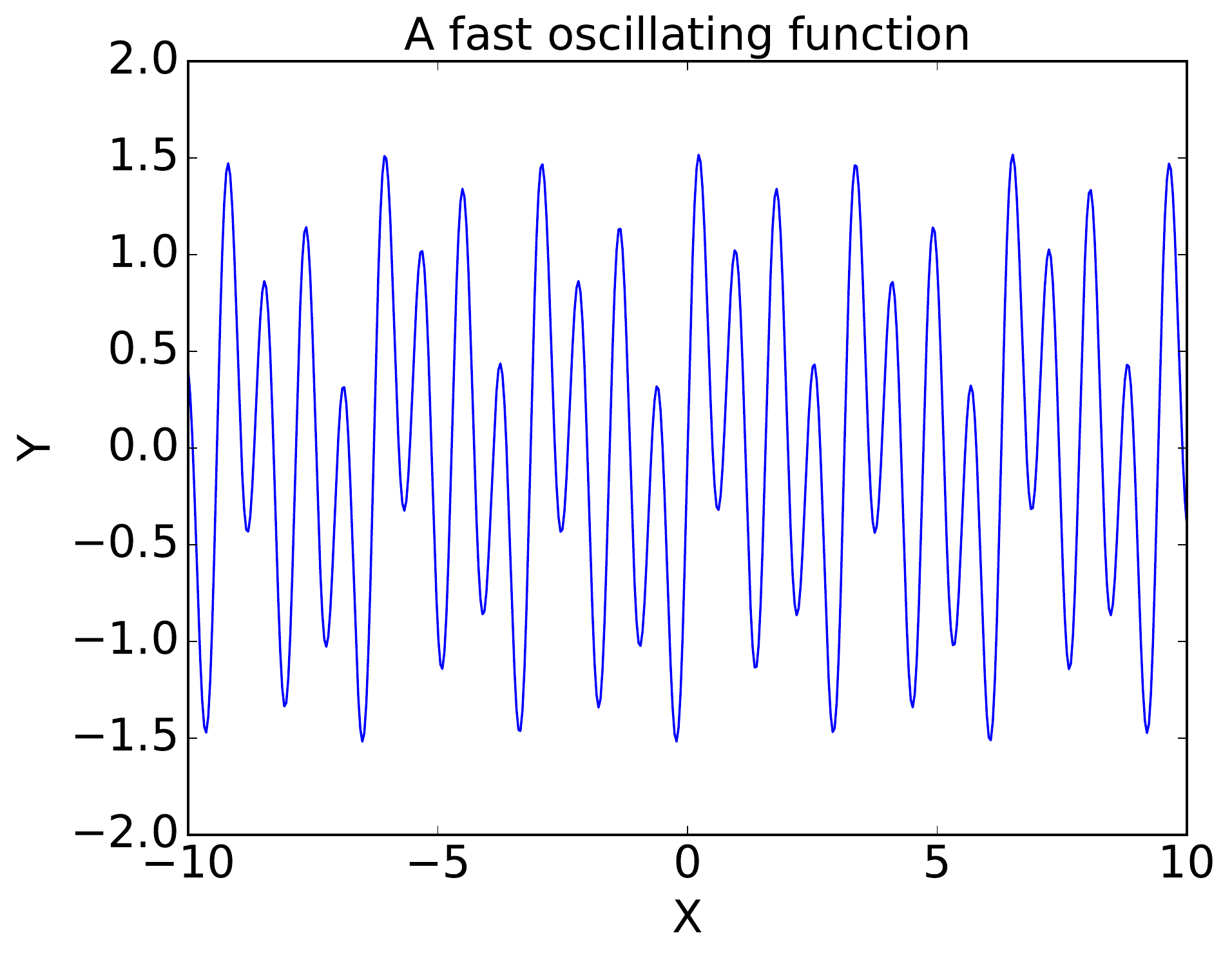}
	\caption{The plot of function~\protect\eqref{eq_fast_oscillation}.\label{fig_fast_oscillation_function}}
\end{figure}

To understand the idea of these algorithms, it is enough to consider $f: \mathbb{R}\rightarrow\mathbb{R}$ defined on a single variable. The simplest gradient descent for finding the minimum of $f$ is the following iterative algorithm: starting from a random number $x_0$ and successively computing $x_{n+1}=x_n-\alpha f'(x_n)$. Gradient based algorithms perform well on functions with non-vanishing gradients almost everywhere and very few local minima, and likely have a poor performance otherwise. For example, the above algorithm would perform very well on a simple function $f(x)=x^2$, but much worse on the following fast oscillating function
\begin{align}
\label{eq_fast_oscillation}
f(x)=&\sin (8x)+0.5\sin (4x)\\&+0.3\sin (2x)+0.1\sin (x)
\end{align}
We plot the above function in Figure \ref{fig_fast_oscillation_function}. It is easy to see we can construct $f(x)=\sum_{i=1}^N a_i \sin (2^i x)$ such that the chance of finding the global minimum is arbitrarily small.

\subsection{Simulated annealing}
Simulated annealing (SA) and its variants stem from the Metropolis-Hastings algorithm. The main idea is constructing a family of probability distribution $p(x,T)$ based on the values of the objective function $f(x)$, with the requirement $p(x,0)>0$ only when $x$ is a global minimum of $f$. Then we repeatedly sample from $p(x,T)$ while slowly decreasing $T$.
In practice, simulated annealing is also an iterative algorithm, i.e. it chooses $x_{n+1}$ based on $x_n$. Since SA uses the Metropolis-Hastings algorithm as a subroutine, there is a non-zero chance to choose $x_{n+1}$ such that $f(x_{n+1})>f(x_n)$. So in principle, SA could escape from local minima, which is an advantage compared to gradient descent. SA also works for functions with discrete variables. As a trade-off, it is likely to be slower compared to gradient descent when $f$ has very few local minima. Moreover, while SA has the mechanism to escape from local minima, in practice it could work poorly on functions with many local minima and high barriers between them, e.g. the Function~\eqref{eq_fast_oscillation}.

\subsection{Genetic algorithms and beyond}
In this subsection we will assume $f$ has the form $f: \mathbb{R}^N\rightarrow\mathbb{R}$. A common feature in all versions of genetic algorithms (GA) is that they maintain a population of solutions $\{ \vec{x}_i, 1\leq i \leq M\}$, where $\vec{x}_i=(x_{i1},\ldots,x_{iN})$. For the first generation, a number of $M'>M$ solutions is randomly generated, then we pick the $\vec{x}_i$ with the $M$ smallest $f(\vec{x}_i)$ as the population. To generate new potential solutions for new generations, several different operations are introduced. In the original genetic algorithm, the two such operations are crossover and mutation. The effect of the mutation operation on a solution $\vec{x}$ is
\begin{align*}
(x_{1},\ldots,x_j,\ldots, x_{N})\rightarrow (x_{1},\ldots,x'_j,\ldots, x_{N}) \ ,
\end{align*}
where $x'_j$ is a random number.
The crossover operation acts on two solutions $\vec{x}$ and $\vec{y}$
\begin{align*}
(\vec{x},\vec{y})\rightarrow (x_{1},\ldots,x_j,y_{j+1},\ldots, y_{N}) \ ,
\end{align*}
where the position $j$ is picked randomly.
Then we can use these two operations to generate $M''$ new test solutions from the first generation, combine them with the $M$ old solutions and pick the top $M$ solutions as the population of the second generation. Later generations can be obtained by repeating these steps.

To illustrate the advantage of the (original) genetic algorithm, we can consider the following objective function $f$
\begin{align*}
f(\vec{x})=\sum_j f_j(x_j) \ .
\end{align*} 
In this case, if $f(\vec{x})$ is (relatively) small, then either $\sum_{j=1}^k f_j(x_j)$ or $\sum_{j=k+1}^N f_j(x_j)$ is (relatively) small. Thus the crossover operations serve as non-local jumps, while the mutation operations help to find local minimum. 
However, in general, it is not clear for what kind of function $f$ the inclusion of the crossover operations could provide an advantage.
It is easy to construct counter-examples such that the crossover operations deteriorate the performance, such as
\begin{align*}
f(\vec{x})=f(\vec{x}_a,\vec{x}_b)=\|\vec{x}_a-\vec{x}_b\| \ ,
\end{align*}
where $\vec{x}_a,\vec{x}_b$ has equal dimension, and $\|\cdot \|$ is the Euclidean norm. Clearly, in most cases, the crossover of two good solutions will only produce inferior new solutions.

It turns out that the most important feature of genetic algorithms is the use of a population. In comparison, other optimization methods we mentioned previously only keep track of the last test solution.
If we are willing to believe that good solutions of the function $f$ have a certain structure (thus partially dropping the black-box requirement of $f$), it is possible that we can identify this structure from the solutions in the population, and then generate new test solutions. This idea has led to the so-called probabilistic model building genetic algorithm (PMBGA) and its variants~\cite{pelikan2002survey,pelikan2005bayesian}. The optimization algorithm we introduced in the main text is also closely related to this idea.

\begin{figure}
	\begin{subfigure}{0.4\textwidth}
	\centering
	\includegraphics[width=\linewidth]{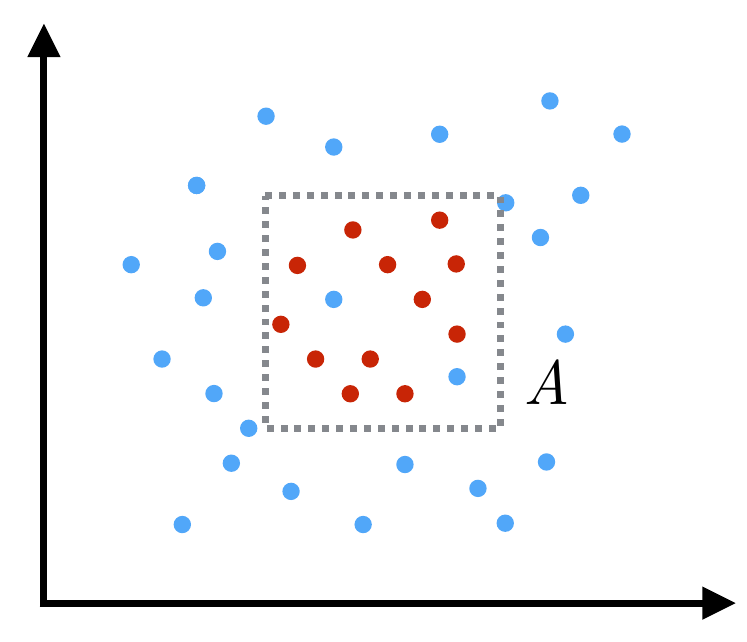}
	\subcaption{Correct hypothesis allows us to sample from a smaller region. (Red points correspond to smaller $f(x,y)$)\label{fig_optimization_hypo}}
	\end{subfigure}
	
	\begin{subfigure}{0.4\textwidth}
		\centering
		\includegraphics[width=\linewidth]{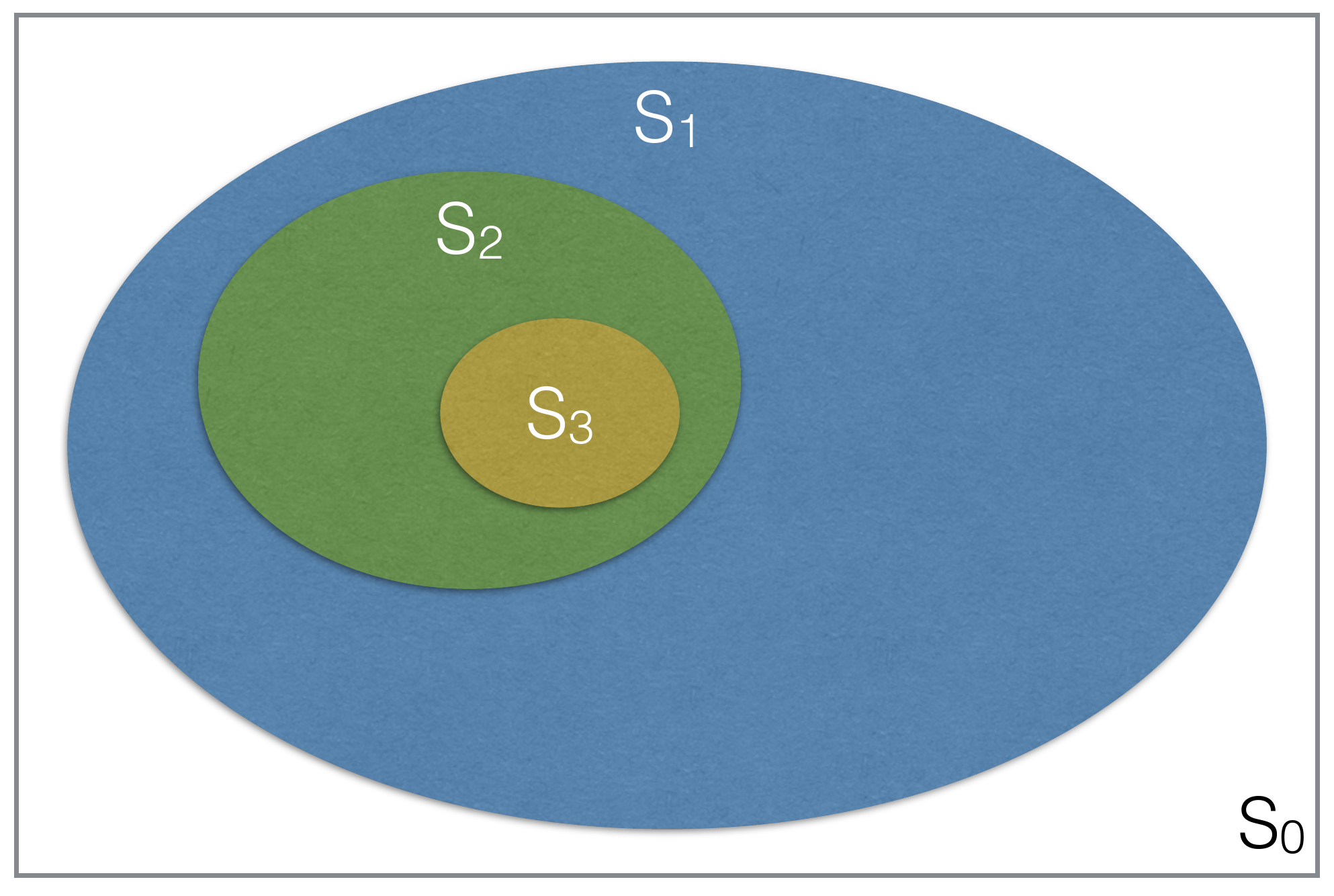}
		\subcaption{Concatenating the operation performed in Figure (a) allows us to sample from sets $S_i$ with better and better solutions.\label{fig_optimization_hierachy}}
	\end{subfigure}
	
	\caption{These two figures can be viewed as an outline of our algorithm. Figure (a) demonstrates that if we can model the distribution correctly, then we will be able to sample from good solutions more efficiently. Figure (b) illustrates the idea of concatenating the step performed in Figure (a) in order to achieve an exponential speedup compared to random search. \label{fig_opti_example}}
\end{figure}
Instead of going through the details of these algorithms, we will explain the idea using a simple example, as illustrated in Fig~\ref{fig_opti_example}. Suppose that we want to minimize a function $f(x,y)$ with two variables which defined on a finite region of $\mathbf{R}^2$, and prior knowledge of $f$ allows us to make the hypothesis $h$ that all points $\{(x,y)\}$ with values $f(x,y)<M$ live in a certain region $A$ (e.g. the rectangular in Fig~\ref{fig_optimization_hypo}). By sampling random points from the domain of the function, we can verify or refute the hypothesis $h$. For simplicity, we assume $h$ is satisfied for all sampled points and $N$ of them is inside the region, then the opposite hypothesis ``an $\alpha$ fraction of points $\{(x,y)\}$ with values $f(x,y)<M$ live outside the region $A$'' will give the observed data a likelihood of $(1-\alpha)^N$. Thus, we can just optimize $f$ over the region $A$ by ignoring a very small fraction of the good solutions. It is easy to see that we can iterate this process, as long as we can formulate a small number of hypothesis such that one of them will describe the good solutions correctly. Our algorithm in the main text resembles this toy example. However, for functions in high dimension and sophisticated generative models such as RNNs, it is hard to give a mathematical justification like in the above example.

It is natural to concatenate the above process (see Figure~\ref{fig_optimization_hierachy}). Let $S_0$ be the domain of $f$, and $S_1$ be the points in region $A$. By sampling enough points from $S_1$, we might be able to build a model and sample from a even smaller set $S_2$ with the good solutions (e.g. find a region $B\subset A$). This way we will introduce a series of sets $\{S_i\}_{i\leq K}$ that we can sample from.
Assuming the order of these subsets satisfies $|S_{i+1}|<\frac{1}{2}|S_i|$, then in the ideal scenario the above iterative algorithm would provide an exponential speedup with respect to $K$.
However, it is worth pointing out that automatically building a model from a data set is, in general, a difficult task (if possible at all).

As another concrete example, we can consider the objective function~\eqref{eq_fast_oscillation} and a routine which looks for the periodicity of the data and then generates new test solutions accordingly.
After we go through multiple generations, it is likely that the population would converge to the correct periodic subset that has the minimum $f(x)$.

\subsection{Summary}

As seen in the discussion above, each of these optimization methods has its strong and weak points. Thus different methods are chosen depending on the prior knowledge we have on the concrete problems. It should be emphasized that we should not consider these methods as in a pure competition; instead, they can be used in complement with each other. For example, stochastic gradient Langevin dynamics (SGLD)~\cite{welling2011bayesian} can be viewed as a combination of gradient descent and annealing, 
and in~\cite{pelikan2006searching}, it is mentioned that inclusion of the deterministic hill climber (discrete version of gradient descent) can lead to a substantial speedup in the PMBGA.


\section{Machine Learning}
\label{sec:ml}
This section will give a brief overview over the subfield of machine learning known as supervised learning and introduce a model for time-series data, known as Recurrent Neural Networks (RNN).
Furthermore, some aspects of the optimization of this class of models will be elaborated on.
\subsection{Supervised Learning}
The field of machine learning can be divided into three main subfields: supervised learning, unsupervised learning and reinforcement learning. These branches differ from each other by the way in which the respective models obtain information about the utility of their generated outputs.

In the case of supervised learning, it is assumed that for every input that a model shall be trained on, a "supervisor" provides a target, corresponding to the desired output of the model for the given input. These pairs of inputs and desired outputs are then used to make the model learn the general mapping between input and output.

More formally and from a Bayesian perspective, one assumes to have a dataset $D$ of size $N$, consisting of several tuples of i.i.d. observations $x \in \mathbb{C}^l$ and corresponding targets $y \in \mathbb{C}^k$, such that \[D=\{(x_i,y_i)\rvert_{i=1}^N \}\]
where $x_i$ and $y_i$ are instances of two random variables $X$ and $Y$ respectively. These random variables are assumed to be distributed according to some unknown probability distribution $p_{Gen}$, the so-called data-generating distribution, \[X,Y \sim p_{Gen}(X,Y).\]
The goal of any supervised learning method now is to approximate the conditional distribution $p_{Gen}(Y|X)$ in a way that allows for evaluation in some new observation $x_* \notin \{x_i\}\rvert_{i=1}^N$.
Since $p_{Gen}$ is not available, one resorts to fitting the empirical distribution $p_{Emp}$ given by $D$ as surrogate problem.

A typical way of deriving a concrete optimization-problem from this is to make an assumption regarding the form of $p_{Gen}$ and treating the model at hand as a distribution $p_M(Y|X,\Theta)$ of this kind, parametrized by the parameters of the model $ \Theta $ that are also often called the \emph{weights} of the model. Now, the fitting of the model can be perceived as a maximum-likelihood problem and hence the supervised learning problem can be formulated as 
\[ \max_{\Theta} \mathcal{L}(\Theta|D) = \max_{\Theta} \, \prod_i p_M(y_i|x_i,\Theta)\text{,} \]
making use of the i.i.d.-assumption. A commonly employed trick to obtain a more benign optimization problem is to instead optimize the \emph{negative log-likelihood}. As the logarithm is a monotonic function, this transformation does not change the location of the optimum in the error landscape, but turns the product of probabilities into a sum over the tuples in $D$. This step then yields a minimization problem, given by
\[ \min_{\Theta} \, -\frac{1}{N} \sum_i \log p_M(y_i|x_i,\Theta) \]
which is also called \emph{empirical risk minimization} (ERM).
These statements of the problem can now be tackled with the optimization methods appropriate for the given model. In the case of the RNN, gradient-based optimization is the state-of-the-art approach and will be explained in Section \ref{opti_rnn}.
 
While it is obvious that fitting a model with respect to $p_{Emp}$ is identical to fitting it to $p_{Gen}$ as long as every tuple in $D$ is only considered once, this is not necessarily true anymore when considering each tuple multiple times. This however is needed by many models in order to fit their parameters to a satisfying degree. In order to prevent the model from learning characteristics of the empirical distribution that are not present in the data-generating distribution, a phenomenon commonly known as \emph{over-fitting}, often some form of regularization is applied. This may be done by punishing too large parameter values, stopping the training after performance starts to decrease on some hold-out data set or by averaging over multiple models. Note that in the Bayesian picture some penalty-terms can be perceived as the logarithm of a prior distribution over $\Theta$, hence turning the optimization problem into finding the \emph{maximum a-posteriori} parameters.
 
\subsection{Recurrent Neural Networks}
\label{RNN}
In this section, the Recurrent Neural Network model will be introduced. We will start with an introduction of the standard version of the model and based upon this, explain the advanced version of the model employed in this work in a second step.
\subsubsection{The Standard RNN Model}
In many areas of application, the data can be perceived as, often non-Markovian, discrete time-series data, such that an observation $x_t \in \mathbb{R}^l$ at some time $t$ depends on the previous observations $x_{t-1},\dots,x_{1}$ or with respect to the framework introduced above,
\[X_t \sim p(X_t| X_{t-1},\dots,X_1) \text{.}\]
While Markov Chains have been the state-of-the-art approach for this kind of data during the last decades, with the recent rise of artificial neural networks, RNNs~\cite{williams1989, werbos1990} have also gained momentum and are now generally considered to be the most potent method.

\begin{figure}
\centering
\includegraphics[scale=0.5]{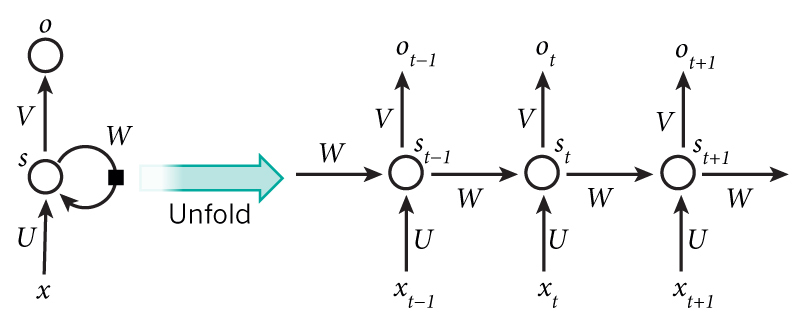}
\caption{The standard model of a Recurrent Neural Network shown for three time-steps.
\label{vanilla_rnn}}
\end{figure}
A RNN is defined by the two non-linear maps $s_t:\mathbb{R}^l \rightarrow\mathbb{R}^h$ and $o_t:\mathbb{R}^h \rightarrow\mathbb{R}^o$ given by \begin{equation}
\begin{aligned}
	s_t &= f_s(Ux_t + Ws_{t-1} + b_s) \\
	o_t &= f_o(Vs_t + b_o) \text{,}
\end{aligned}
\end{equation}
where $U \in \mathbb{R}^{h \times l}$, $W \in \mathbb{R}^{h \times h}$, $V \in \mathbb{R}^{o \times h}$, $b_s \in \mathbb{R}^{1 \times h}$, $b_o \in \mathbb{R}^{1 \times o}$ and the trainable parameters of the models are constituted by $\Theta=\{U,V,W,b_s,b_o\}$. The non-linear function $f_s$ is often chosen to be $tanh$, the rectifier-function given by \[rect(x)=\max(0,x)\] or the sigmoid-function given by \[sigm(x)=\frac{1}{1+e^{-x}} \text{.}\]
The function $f_o$ must be chosen according to the distribution that is to be approximated by the model. For the case of a multinoulli distribution as assumed in this work, the corresponding function would be the $softmax$, defined as \[softmax(x)_j=\frac{e^{x_j}}{\sum_k e^{x_k}} \text{,}\]
the superscripts in this case denoting the single elements of the vector $x$.

The intuition behind this simple model is that it combines its information about the input at a given time step with a memory of the previous inputs, referred to as \emph{state} of the network. The precise nature of this combination and the state depends on the weight matrices $U$ and $V$ and the bias-vector $b_s$. The combined information is then used as input of the chosen non-linear function $f_h$ to generate the next state. From this state, the output $o_t$ is then computed as defined by $W$, $b_o$ and $f_o$.
The effect of an RNN acting on the sequence $\{x_t\}$ is illustrated in Figure \ref{vanilla_rnn}.
\begin{figure}
	\centering
	\includegraphics[scale=0.5]{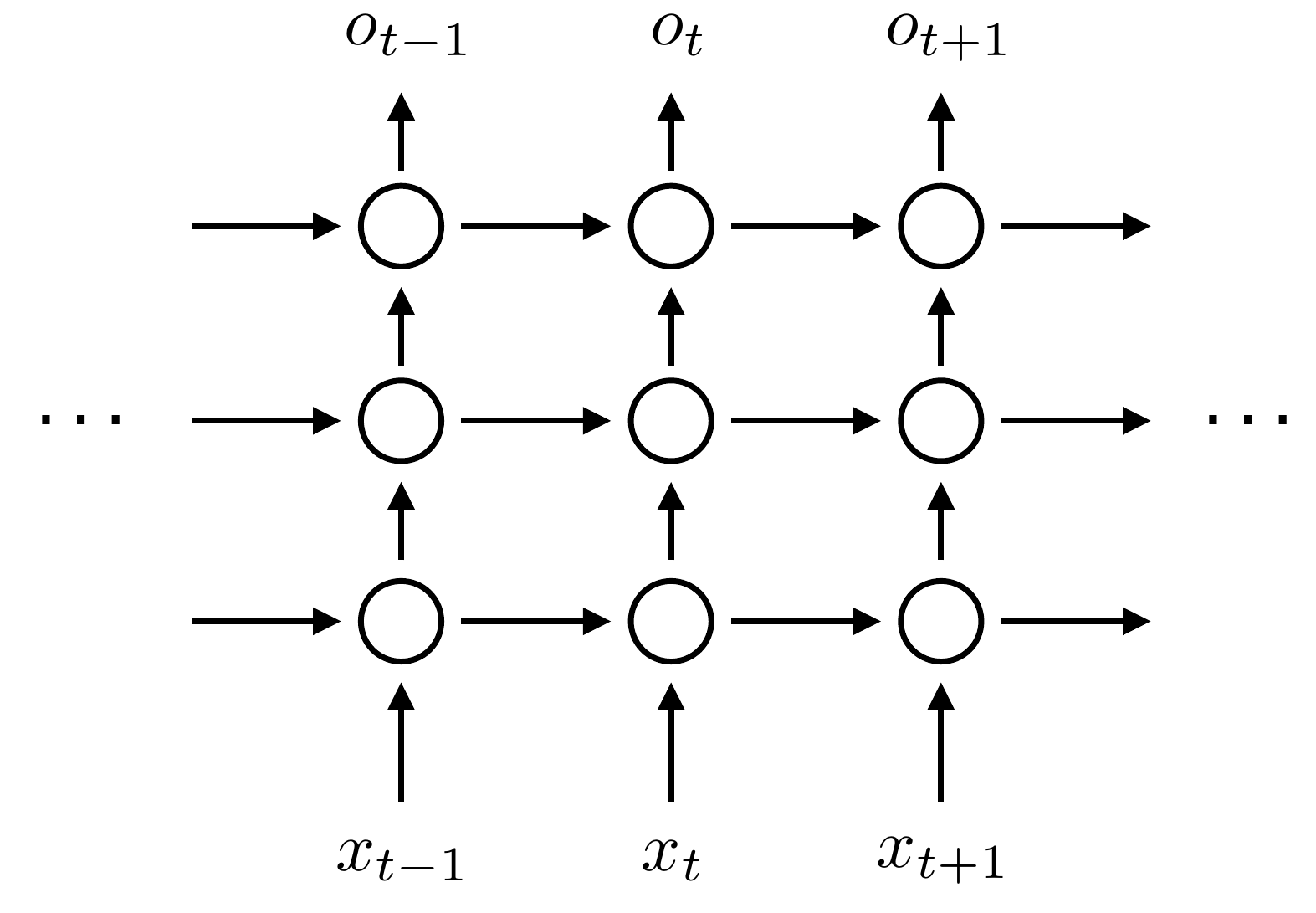}
	\caption{An illustration of an RNN with 3 hidden layers.
		\label{fig_multi_layer_rnn}}
\end{figure}

From the above explanation, it is clear that the power of the model depends strongly on the size of the hidden state $h$. It should however also be noted that another effective way of increasing the expressive power of an RNN is to construct a composition of multiple functions of the form of $s_t$, see Figure~\ref{fig_multi_layer_rnn}. 
In the machine learning terminology, the respective functions are called the \emph{layers} of an artificial neural network and the number of composed functions is referred to as the depth of a network. The layers between the input and the output are referred to as \emph{hidden layers}. The common intuitive reasoning behind stacking multiple layers is that it will allow the network to learn a hierarchy of concepts, called \emph{features}, from the initial input data. Thereby, the features are assumed to be of increasing complexity with every layer, as they are based on a linear combination of the features learned by the layer below. Apart from this intuitive reasoning, also more rigorous work on the benefits of using at least one hidden layer between input and output can be found in the literature\cite{bengio2007scaling, hornik1989, hornik1991}. This ansatz of increasing the power of neural network models via deepening their architecture is publicly known as \emph{Deep Learning} and has led to a drastic increase in success of machine learning methods during the last decade.
However, having a composition of many state-computing functions of similar size can slow down the optimization process. This is why, when forming such a composition, each pair of functions is often connected via a simple linear projection from the space of the state of the earlier function onto some lower-dimensional space that is then used by the following function.
Note that while all the above claims seem natural and lead to a good enough performance for our paper, more benchmarking is needed to really confirm them.

Now, in the case of supervised learning, one assumes to be in possession of a set of time-series $x_1,\dots,x_n$ that shall be used to let the RNN learn to predict series of this kind. The natural way of doing this is to define the pairs $(x_i,y_i) \coloneqq (x_t,x_{t+1})$. While in principle the model is capable of taking into account all previous time steps, in practice it shows that optimization is only feasible for a relatively short number of steps. This is mainly due to the fact that the gradients that are needed to optimize the parameters of an RNN tend to grow to infinity or zero for higher numbers of steps. This will be discussed more in-depth below.
\subsubsection{Long Short-Term Memory Networks}
In order to improve upon the standard RNN, Hochreiter et.\ al.\ introduced the Long Short-Term Memory network (LSTM)~\cite{hochreiter1997}, which provides a different way of computing the state of an RNN. Hence the following set of equations can be perceived as a replacement for $s_t$ from the previous section. The main advantage of the approach is that it drastically mitigates the problem of unstable gradients by construction. It is defined by the following set of equations,
\begin{equation}\label{eq:lstm}
\begin{aligned}
	i_t &= sigm(U^ix_t + W^is_{t-1} + b^i) \\
	f_t &= sigm(U^fx_t + W^fs_{t-1} + b^f) \\
	o_t &= sigm(U^ox_t + W^os_{t-1} + b^o) \\
	\tilde{c}_t &= tanh(U^{\tilde{c}}x_t+W^{\tilde{c}}s_{t-1} + b^{\tilde{c}}) \\
	c_t &= c_{t-1} * f_t + \tilde{c}_t * i_t \\
	s_t &= tanh(c_t) * o_t
\end{aligned}
\end{equation}
where again $x_t$ is the input at time step $t$, $s_{t-1}$ is the previous state of the network and $c_t$ is the state of the cell. $U^i, U^f, U^o, U^{\tilde{c}} \in \mathbb{R}^{h \times l}$, while $W^i, W^f, W^o, W^{\tilde{c}} \in \mathbb{R}^{h \times h}$, $b^i, b^f, b^o, b^{\tilde{c}} \in \mathbb{R}^{1 \times h}$ and $*$ denotes the element-wise multiplication.

\begin{figure}
\centering
\includegraphics[scale=0.7]{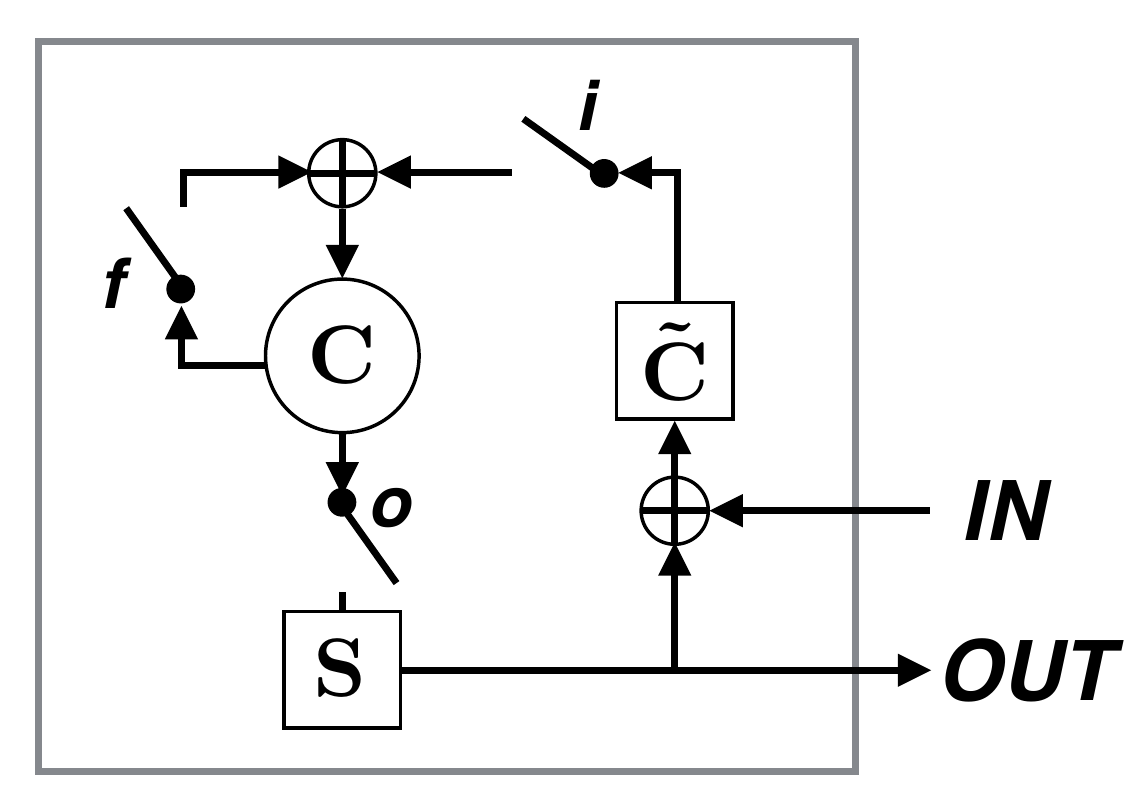}
\caption{The Long Short-Term Memory model illustrated in a schematic way.
	In addition to the diagram above, the input gate $\mathbf{i}$, the forget gate $\mathbf{f}$ and the output gate $\mathbf{o}$ all depend on the current input $x_t$ and previous state of the network $s_{t-1}$, as described in~\protect\eqref{eq:lstm}.
\label{lstm}}
\end{figure}
As it can be seen from the equations, the way in which an LSTM computes the state is a bit more involved. If needed, it may however just be treated as a black box and can be stacked just in the same manner as it was described for the plain RNN model. The general idea of an LSTM is to give the model a higher degree of control over the information that is propagated from one time step to the next. This is achieved by making use of so-called \emph{gates} that control the information flow to and from the network and cell state. These gates, by taking into account the previous  state and the new input, output vectors of values in $[0,1]$ that determine how much information they let through. In the equations given above, $i_t$ is called the \emph{input gate}, $f_t$ is referred to as the \emph{forget gate} and $o_t$ denotes the \emph{output gate}. Now, the mechanism works as follows:
\begin{itemize}
\item For a given time step $t$, the new input and previous network state are processed by $\tilde{c}_t$ like for the standard RNN and the output values are squashed to the interval $[-1,1]$ to yield candidate values for the next cell state.
\item The input gate $i_t$ determines how to manipulate the information flow from the candidate cell state. Likewise, the forget gate $f_t$ determines how to affect the information flow form the old cell state. The gated previous cell state and the gated input are then added to form the new cell state $c_t$.
\item Finally, the output gate $o_t$ determines what to output from the new cell state. The new cell state is then also projected onto the interval $[-1,1]$ and put through the output gate to become the network state.
\end{itemize}
The whole process is shown in Figure \ref{lstm}.

Naturally, there exists a plethora of possibilities to adapt the normal LSTM as explained above. One important enhancement is commonly referred to as \emph{peepholes}, which allows the gates to incorporate the cell state via an extra term in the sum, in addition to the input and the network state. One other popular possibility introduced in~\cite{sak2014} is the use of projection layers between different time steps of LSTM. In this case, we replace $s_{t-1}$ by $r_{t-1}$ in the equations for $i_t, f_t, o_t$ and $\tilde{c}_t$ and add the simple equation
\begin{equation}
r_t = W^ps_t
\end{equation} 
where $W^p \in \mathcal{R}^{k \times h}$ is the projection matrix. In this work, we have made use of both of these extensions of the normal LSTM.
For an exhaustive overview over the known variants of the LSTM, we refer the interested reader to~\cite{greff2015}. 
\subsection{Optimization of RNNs}
\label{opti_rnn}
As the optimization problem described in the beginning of this section can not be solved analytically for the models considered in this work, gradient-based approaches have established themselves as the state of the art. However, in the case of fitting the parameters of neural network models, three main restrictions need to be accounted for:
\begin{enumerate}
\item The number of parameters for neural network models easily exceeds 100,000 and can for larger architectures go up to several tens or even hundreds of millions. Hence, computing the Hessian (or its inverse) explicitly is not tractable and so, one is limited to first-order or approximative second-order methods.
\item As the error function that is minimized is only a surrogate error function, its global optimum is not necessarily the optimum of the error function one actually wants to minimize.
\item For many real-world data sets, computing the gradient of the complete sum of the error function over all samples is not feasible. Hence, the sum is normally split up into smaller parts called \emph{mini batches} and these batches are looped over. A complete loop over $D$ is then called an $epoch$.  
\end{enumerate}
These restrictions have led to the rise of an own subfield of machine learning that is concerned with the parallelization of gradient computations in the mini batch case, the approximation of second-order information and the formal justification for the splitting up of the error function. All of the currently available methods are nevertheless extensions of the simplest method for gradient-based optimization known as \emph{steepest gradient descent}: At iteration $i$ in the loop over the batches, the parameters $\Theta$ are updated according to
\[\Theta_{i+1}=\Theta_i- \gamma \frac{\partial \mathcal{E}}{\partial \Theta_i} \]
where $\mathcal{E}(D,\Theta)$ is the respective error function and $\gamma$ is called the \emph{step rate}. The most straight forward natural adaption is to make $\gamma$ depend on the iteration and slowly decrease it over time, following the intuition that smaller steps are beneficial the closer one gets to the respective optimum. In addition to that, many methods employ some kind of momentum term~\cite{sutskever2013} or try to approximate second order information and scale the gradient accordingly~\cite{tieleman2012, kingma2015}.

Besides this, the size of the batches also has an influence on the performance of the respective optimization method. In the extreme case where each batch only consists of one sample, the gradient descent method is known to converge almost surely to an optimum under certain constraints~\cite{saad2009}. As picking individual samples for optimization can be perceived as sampling from the empirical distribution to approximate the overall gradient, this method is called \emph{stochastic gradient descent} (SGD). Using single data points however is computationally inefficient and by definition leads to heavily oscillating optimization, so it is common practice to resort to larger batches. Following the ERM-interpretation, batches $B$ consisting of $S_B$ samples are often used to compute an approximation of the mean gradient over $D$ given by \[ \langle \frac{\partial \mathcal{E}}{\partial \Theta_i} \rangle_{D} \approx \langle \frac{\partial \mathcal{E}}{\partial \Theta_i} \rangle_{B} = \frac{1}{S_B} \sum_{(x,y) \in B} \frac{\partial \mathcal{E}_{(x,y)}}{\partial \Theta_i} \]
where obviously \[\lim_{|B| \rightarrow |D|} \langle \frac{\partial \mathcal{E}}{\partial \Theta_i} \rangle_{B} = \langle \frac{\partial \mathcal{E}}{\partial \Theta_i} \rangle_{D}.\] This interpretation is used, e. g. by the recently proposed algorithm \emph{Adam} which has been shown to yield very good local optima while being very robust with respect to noisy gradients and needing comparatively little adjustment of its parameters. We have employed Adam for fitting the models used in this work.

While the approach to optimizing artificial neural networks is well established, this does not change the fact that the optimization problems posed by them are inherently difficult. It is well known that the error landscape becomes less smooth the more layers one adds to a network. This results in error surfaces with large planes where $\frac{\partial \mathcal{E}}{\partial \Theta} \approx 0$ that are followed by short but very steep cliffs. If the step rate is not adapted correctly, the optimization procedure is very likely to get stuck in one these planes or saddle points and to jump away from an optimum in the vicinity of $\Theta$ if evaluated on one of the cliffs. The phenomena of the frequent occurrence of very large or very small gradients are referred to in the literature as the \emph{exploding gradient} or \emph{vanishing gradient} problem respectively. To get a better understanding of why these problems exist, it is instructive to examine how the gradients for a given model are obtained.

As has been explained above, multi-layer neural network models are a composition of non-linear functions $\mathbb{R}^{i_k} \rightarrow \mathbb{R}^{o_k}:x_{k+1}=f_k(W_kx_k+b_k)$, where $W_k$ is the weight-matrix, $b_k$ the bias-vector, $x_0$ the input data and $x_K$ the final output of the network.
From this definition it is clear that $o_k=i_{k+1}$.
For convenience, we define $y_k\equiv W_kx_k+b_k$.
In order to obtain the gradient for a specific $W_k$ or $b_k$ one must obviously make use of the chain rule, such that 
\begin{align*}
	\frac{\partial \mathcal{E}}{\partial W_k} &= 
	\frac{\partial \mathcal{E}}{\partial x_{k+1}}
	\frac{\partial x_{k+1}}{\partial y_k}
	\frac{\partial y_k}{\partial W_k}\\
	 &= \frac{\partial \mathcal{E}}{\partial x_K} \left(\prod_{j=k+1}^{K-1} \frac{\partial x_{j+1}}{\partial x_j}  \right)
	\frac{\partial x_{k+1}}{\partial y_k}
	\frac{\partial y_{k}}{\partial W_k}
\end{align*}
and 
\begin{align*}
	\frac{\partial \mathcal{E}}{\partial b_k} &=  \frac{\partial \mathcal{E}}{\partial x_{k+1}}
	\frac{\partial x_{k+1}}{\partial y_k}
	\frac{\partial y_k}{\partial b_k}\\
	&= \frac{\partial \mathcal{E}}{\partial x_K} \left(\prod_{j=k+1}^{K-1} \frac{\partial x_{j+1}}{\partial x_j}  \right)
	\frac{\partial x_{k+1}}{\partial y_k}
	\frac{\partial y_{k}}{\partial b_k}
\end{align*}
where $\frac{\partial}{\partial W_k}$ is the shortcut of doing the derivative element-wise:
\begin{align*}
\left[ \frac{\partial}{\partial W_k} \right]_{ab}=\frac{\partial}{\partial [W_k]_{ab}}
\end{align*}
The same convention applies to $\frac{\partial}{\partial b_k}$.
As $\frac{\partial}{\partial W_k}$ and $\frac{\partial}{\partial b_k}$ depend on all the gradients of the later layers, this formulation yields an efficient method of computing the gradients for all layers by starting with the uppermost layer and then descending in the network, always reusing the gradients already computed. Together with the fact that many of the commonly used non-linearities have an easy closed-form expression of the first derivative, this allows for fully automatic computation of the gradients as it is done in every major deep learning framework. This dynamic programming method of computing the gradients is known in the literature as \emph{Back-Propagation}.
The vanishing (exploding) gradient problem arises because of the product $\prod_{j=k+1}^{K-1} \frac{\partial x_{j+1}}{\partial x_j}$ in the above equations.
For example, if one of the $\frac{\partial x_{j+1}}{\partial x_j}\approx 0$ in the product, then likely we have $\frac{\partial \mathcal{E}}{\partial W_k} \approx 0$, which leads to an ineffective gradient descent.
Similarly, if many of the terms $\frac{\partial x_{j+1}}{\partial x_j}$ have large norms, then there is a possibility that $\frac{\partial \mathcal{E}}{\partial W_k}$ becomes too large, which often causes the optimization method to jump out of a local optimum.
\begin{figure}
	\centering
	\includegraphics[width=0.5\textwidth]{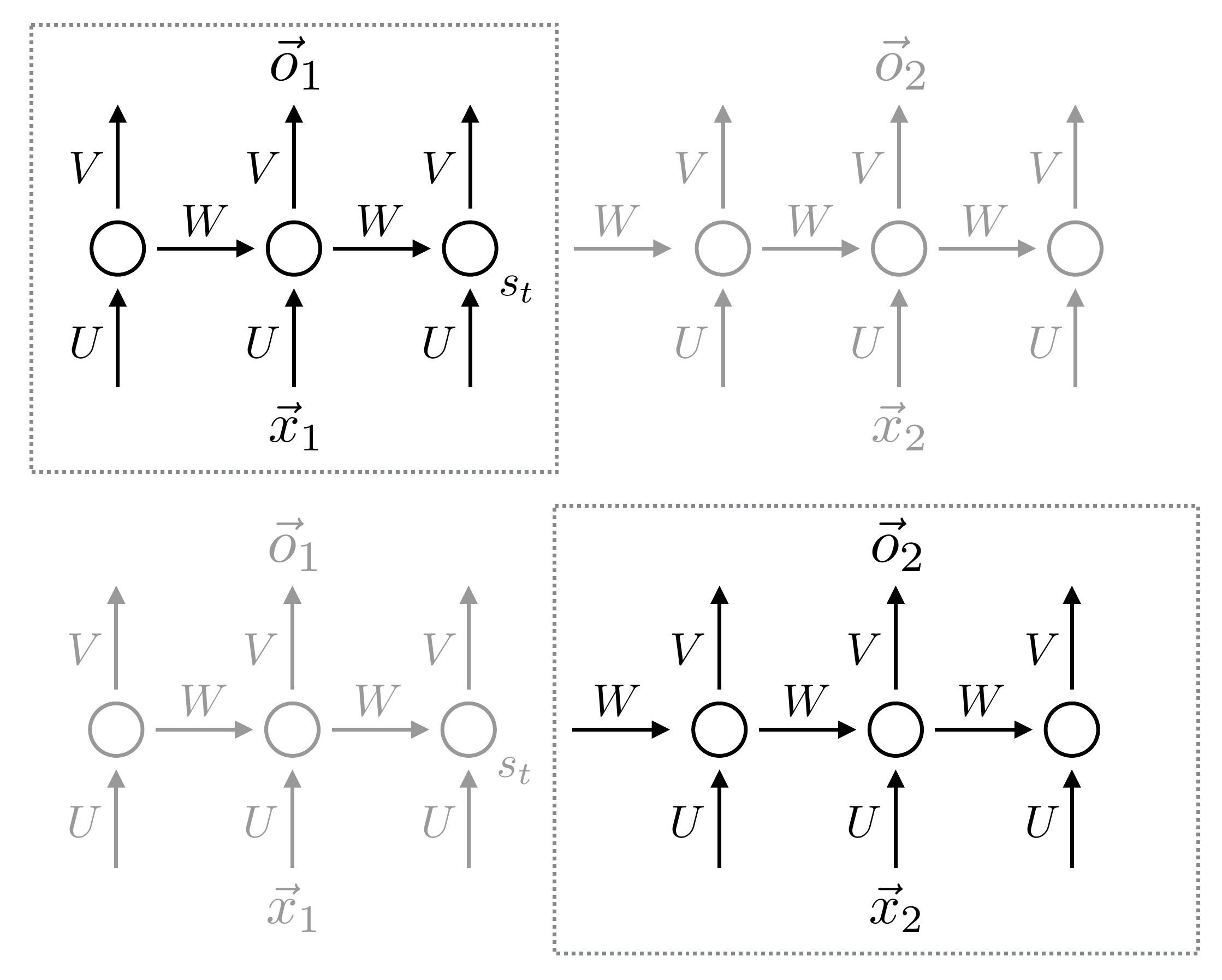}
	\caption{An illustration of how we truncate the gradient computation for long sequences.
		Here we divide the sequences into two halves.
		As the first step, we compute the gradient of the error function $\mathcal{E}(\vec{x}_1,\vec{o}_1)$ with respect to the parameters $U,V,W$, while ignoring the other half of the network.
		In the second step, we compute the gradient of $\mathcal{E}(\vec{x}_2,\vec{o}_2)$, while treating the final state of the network $s_t$ of the first half as a constant.
		The final gradients are approximated by the sums of these two constituents.
		Thus, we are able to avoid the instability of computing gradients, but still capture the correlation between two halves, since we feed the final network state $s_t$ into the second half.
		\label{fig_rnn_trunc}}
\end{figure}

In the case of an RNN as defined in Section \ref{RNN}, the above generic equations for the derivative become a little more involved, as in addition to the term for possibly multiple stacked layers, a term accounting for states of previous times has to be added.
Nevertheless, at the heart of the problem, it is still about computing derivatives of composite functions.
This slightly more involved back-propagation method is known as \emph{Back-Propagation through Time} and can also be fully automatized.
Similar to the multi-layer neural network models mentioned above, the gradient computation of RNNs also has these instability issues.
As can be seen from Figure~\ref{vanilla_rnn}, the same matrix $W$ is used in all time step of an RNN.
Thus, a tiny change of $W$ could affect the output $o_t$ drastically when the time step $t$ gets big.
In other words, the derivative of the error function $\mathcal{E}$ with respect to $W$ could again become very large or very small in certain situations.
To deal with this issue, we could truncate the number of time steps during the computation, as described in Figure~\ref{fig_rnn_trunc}.
More discussion on this topic can be found in Section 3.2 of~\cite{Graves2013}.

\section{Technical Aspects}
For the implementation of this work, we have made use of Python with the numerical libraries NumPy, SciPy and TensorFlow~\cite{numpy, scipy, tensorflow}. All experiments were run on single workstations with up to 8 threads and a GeForce Titan X. The runtime of the experiments varied, depending on the optimization parameters, from a few hours to several days.

\end{document}